\documentclass[a4paper,12pt,twoside]{article} % 12pt,11pt,10pt

\usepackage{graphicx}
\usepackage[textwidth=8em,textsize=small]{todonotes}

\usepackage{amsmath}
\usepackage{natbib}
\usepackage{siunitx}
\usepackage{amsfonts}
\usepackage{mathtools}
\usepackage{amssymb}
\usepackage[retainorgcmds]{IEEEtrantools}
\usepackage[T1]{fontenc}
\usepackage{array,collcell}
\newcommand{\myverb}[1]{\ttfamily\detokenize{#1}}
\usepackage{graphicx}
\usepackage{amsmath,amssymb}

\usepackage{hyperref}
\usepackage{etoolbox}

\makeatletter
\patchcmd{\@makecaption}
  {\parbox}
  {\advance\@tempdima-\fontdimen2} % decrease the width!
  {}{}
\makeatother

\usepackage[caption = false]{subfig}
\usepackage{ntheorem}

\newtheorem*{theorem-non}{Remark}
\newtheorem*{pro}{Proof}
\newtheorem{coro}{Corollary}
\newtheorem{thm}{Theorem}
\newtheorem{lem}{Lemma}

\begin{document}
%\color{blue}
%pagestyle{plain}
{\Large
\vspace*{0.1cm}
\textbf{Testing for equal correlation matrices \\ with application to paired gene expression data
}}
\vspace{0.2cm}
\\
\vspace{0.1cm} Adria Caballe$^{a,b}$, Natalia Bochkina$^{a,c}$, Claus Mayer$^b$, Ioannis Papastathopoulos$^{a,c}$\\
$^a$ University of Edinburgh \& Maxwell Institute, EH9 3FD, Scotland, UK\\
$^b$ Biomathematics \& Statistics Scotland, Scotland, UK\\
$^c$  The Alan Turing Institute, Cambridge, UK

\begin{abstract}
% In this paper
 We present a novel method for testing the hypothesis of equality of two correlation
    matrices using paired high-dimensional datasets.\ We consider test statistics based on the
    average of squares, maximum  and sum of exceedances of Fisher transform sample correlations
    and we derive approximate null distributions using asymptotic and non-parametric distributions.\
    Theoretical results on the power of the tests %under different alternatives
    are  presented and backed up by a range of simulation experiments.\
    We apply the methodology to a %high-dimensional
    case study of colorectal tumor  gene expression data with the aim of discovering biological pathway lists of genes
    that present significantly different correlation matrices on healthy
    and tumor samples.\ We find strong evidence for a large part of the
    pathway lists correlation matrices to change among the two
    medical conditions.%, especially when using the average of squares and
%    sum of exceedances tests.
\end{abstract}

\textbf{keywords} correlation matrix, Fisher transform, hypothesis testing,  high-dimensional data, extreme values, dependent data, gene expression  %, exceedance,




%\end{keyword}

%\color{blue}
%%%%%%%%%%%%%%%%%%%%%%%%%%%%%%%%%%%%%%%%%%%%%%%%%%%%%%%%%%%%
%%%%%%%%%%%%%%%%%%%%%%%%%%%%%%%%%%%%%%%%%%%%%%%%%%%%%%%%%%%%
%\vspace{-1cm}
\section{Introduction}

In recent years, the improvements in technology have made it possible to
collect and store reliable information for a large number of genes,
metabolomics or proteins, among others, on an organism in a single
sample.\ This typically generates datasets where the number of variables $p$ is
much larger than the number of observations $n$.\ 
Statistical techniques
that deal with this type of data, commonly known as high-dimensional data,
with the purpose of answering biological questions, are well studied in the
literature \citep{Buhlmann2010, Sanchez2008}.\ One of the main challenges
relates to understanding how the genes function in a biological process and
how they interact between each others in a cell.\ In this regard,
measuring and assessing  variations of gene interactions on the presence of an illness process such as cancer
is important to biologists as part of discerning the gene regulatory mechanisms that control the disease.

A statistical technique that is widely used to measure interaction between
pairs of genes from data is given by the Pearson correlation, which quantifies
the strength of the linear dependence between two random variables.\
The main hypothesis testing (HT) problem we study in this chapter assesses the evidence of 
 equality of two correlation matrices $R_1 = [r_{ij}^{(1)}]$ and $R_2 = [r_{ij}^{(2)}]$ that correspond to genomic data $Y^{(1)}$ and $Y^{(2)}$ measured in 
two different conditions (e.g, healthy and tumor tissues), 
\[
 \text{H}_0: R_1=R_2\,\,\,\,\, \text{vs}\,\,\,\,\, \text{H}_1: R_1\neq R_2
\]

The likelihood ratio test statistic for testing equality of correlation matrices
\citep{Kullback1967}, when the vectors $Y^{(1)}$ and $Y^{(2)}$ are Gaussian, depends 
on the determinant of the two sample matrices and is not well defined when the sample size $n$ is
less than the dimension $p$.\ However, datasets that arise from biological
experiments are frequently high-dimensional, with $p\gg n$.\ There are two
main directions that address this hypothesis testing problem for high-dimensional
data in the literature.\ The first is based on sum of squares statistics, see \cite{Schott2007}
and \cite{Li2012a}, who use the Frobenius norm as a distance measure to
compare the two sample correlation matrices.\ The second is based on extreme
value statistics, see \cite{Cai2013a}, who derive the asymptotic null distribution of 
the maximum of the square of sample correlation coefficient differences.

To the best of our knowledge, the tests considered so far in the literature
are applicable when the random vectors $Y^{(1)}$ and $Y^{(2)}$ are independent.\ Here we study the
implications of using the sample correlation matrices when the two datasets
are dependent, particularly when they come from paired observations, in which case the
cross-correlation is not zero.\ %$R_{XY}\neq 0$ (e.g. paired data).
We propose three different tests which apply to paired data, and that are based
on the average, maximum and threshold exceedances of the elementwise
correlation differences.
%In all of them, an initial transformation to the sample correlation estimator is applied in order
%to improve the sampling distribution. The transformation chosen is the
%Fisher transformation \citep{Fisher1921}. The three tests are subsequently
%based on the average, maximum and threshold exceedances of the
%elementwise differences.

%for very sparse alternatives. If the alternatives are not (as extreme,
%{\ccom don't understand}), for a reasonable selection of the
%threshold, the exceedance test overperforms the power of the other two
%tests.
%We provide an expression for the limiting distribution of the test statistic under $H_0$
%assuming independence between sample correlation coefficients. This provides
%computational fast approaches close to the current methods in the literature.
%However, we show that these only present a good representation of the null
%distribution for large $n$ (order of hundreds) and very sparse correlation matrices.
%To overcome problems when these conditions do not hold we introduce adjustments
%on the given parametric distribution using permuted datasets.   We finally compare them
%with a non-parametric distribution that captures the dependence structure between sample
%correlation coefficients but  does not assume any parametric distribution for the test
%statistic under $H_0$.

The proposed methodology is motivated by a genomic data set \citep{Hinoue2012} that
contains the gene expression information of approximately  $p\approx \num{25e3}$ genes in
two different samples, from the same $n=25$ patients, corresponding to two different medical conditions.\
These are the gene expression of a tumor cell and its adjacent normal tissue.\ 
The gene pairwise correlation is a reasonable measure to understand the relationship between genes in a 
biological process, so our purpose in the analysis of these data is to assess whether the correlation matrix 
varies or not when going from a healthy to a tumor state.\
Even though the complete $p\times p$ correlation matrix is expected to change
considerably, testing the equality of linear dependence structures for  subgroups of the $\num{25e3}$ genes
that are known to have functions in a biological process is highly important.\  We test if the genes interact
similarly in the two conditions for $1320$ pathway lists of sizes going from $20$ to $900$ which contain groups
of genes with known biochemical connections.

The article is structured as follows.\ In Section \ref{SEC2} we present the hypothesis testing problem and
we propose several test statistics which are motivated by the type of statistics mentioned in the literature.\ 
In Section \ref{SEC3} we determine approximate distributions of these test statistics
under the null hypothesis and we give  lower bounds  for their asymptotic powers.\
In Section \ref{SEC4} we use simulated data in order to assess the accuracy of the
tests under the null hypothesis and to compare the power of the tests for several types of alternative hypothesis.\
Finally, in Section \ref{SEC5} we consider a case study on genomic data where the proposed methodology is used to
 answer questions that arise from a biological process. We have implemented the methodology presented in this paper
 within the R package  \textbf{ldstatsHD} \citep{caballe2016b}.\

%%%%%%%%%%%%%%%%%%%%%%%%%%%%%%%%%%%%%%%%%%%%%%%%%%%%%%%%%%%%
%%%%%%%%%%%%%%%%%%%%%%%%%%%%%%%%%%%%%%%%%%%%%%%%%%%%%%%%%%%%
%\s%ection{Material and methods}
\section[Hypothesis testing problem]{Hypothesis testing problem}\label{SEC2}
\subsection{Problem setting and Fisher transformation}
Consider $n$ independent and identically distributed (i.i.d.) $2p$-dimensional random vectors $Y_k = (Y_k^{(1)}, Y_k^{(2)})$, $k=1,\ldots,n$, where $Y^{(1)}$ and $Y^{(2)}$ are associated with population I and  population II, respectively, and that  follow a  standard multivariate normal distribution with correlation $R$, i.e.,
\begin{equation}\label{eq:jointCorHT}
(Y_k^{(1)},Y_k^{(2)}) \stackrel{iid}{\sim} N_{2p}(0,R), \mbox{\hspace{0.6cm}} R = [r_{ij}] = \begin{bmatrix} R_1 & R_{12} \\ R_{12}^\intercal  & R_2  \end{bmatrix},
\end{equation}
where $R_1$ and $R_2$ are the category-specific correlation matrices and the cross-correlation $R_{12}$ is non-zero if the two random vectors $Y^{(1)}$ and $Y^{(2)}$ are linearly dependent.\ We assume, without loss of generality, unit variances and zero mean vector.\  
The main goal of this section is to test whether the correlation matrix $R_1$ is equal to the correlation matrix $R_2$ with hypothesis $H_0 : R_1 = R_2\,\,\, \text{vs}\,\,\, H_1 : R_1\neq R_2$.
We denote the sample correlation matrix by $\hat{R}$, which is determined by $\hat{R}_1 = [\hat{r}^{(1)}_{ij}] = Y^{(1)^\intercal} Y^{(1)}/n$, $\hat{R}_2 = [\hat{r}^{(2)}_{ij}] =Y^{(2)^\intercal} Y^{(2)}/n$ and $\hat{R}_{12} = [\hat{r}^{(12)}_{ij}] = Y^{(1)^\intercal} Y^{(2)}/n$.\
Given the symmetry in the correlation matrices, we consider their lower triangular matrices instead using the same notation with
\begin{equation}\label{standDif2}
M=\{(i,j) \in \{1,\ldots, p\}: i < j\},\quad  m = \text{Card}(M) =  p\,(p-1)/2.
\end{equation}
An approximate pivot for the correlation coefficient is given by the Fisher transformation \citep{Fisher1921}, which is defined by  $g: (-1,1)\mapsto \mathbb{R}$, $g(z) = \log\{(1+z)/(1-z)\}/2$, such that the elementwise Fisher transformation of $\hat{R}_K$,  $K \in \{1,2\}$, weakly converges to a 
multivariate normal distribution
\begin{equation}\label{eq:ztransf}
\hat{U}_K = g(\hat{R}_K) \sqrt{n-3} \sim N(g(R_K)\sqrt{n-3},\Psi_K), \,\,\,\,\, K \in \{1,2\},
\end{equation}
where $\Psi_K= [\psi^{(k)}_{th}]$ is the $m\times m$ correlation matrix between elements in $\hat{U}_K$ as $\psi^{(k)}_{tt} = 1$ for any $t \in M$ and $K \in \{1,2\}$.\

%\begin{equation}\label{eq:ztransf}
%\hat{u}_{ij} = g(\hat{r}_{ij}) \sqrt{n-3} \sim N(g(r_{ij}),1), \mbox{ \hspace{0.6cm} } \hat{U} = \begin{bmatrix} \hat{U}_X & \hat{U}_{XY} \\ %\hat{U}_{XY}^t  & \hat{U}_Y  \end{bmatrix} = [\hat{u}_{ij}].
%\end{equation}

%From now on, given the symmetry in the correlation matrices, we consider the lower triangular matrices of $\hat{U}_X, \hat{U}_Y, \hat{U}_{XY}$ using the same notation.

\subsection{Correlation of sample correlation coefficients}\label{paired}
We assume here and  throughout that $r_{t} < 1$ for any $t \in M$.\ The non-zero dependence structure between the 
two random vectors $Y^{(1)}$ and $Y^{(2)}$  leads to correlation between elements in the estimator $\hat{U} = [\hat{U}_1, \hat{U}_2]$ 
\citep{Elston1975,Steiger1980}, which is found as in eq.~\eqref{eq:ztransf}.\
Take  $s = (h,i)$ and $t = (j, l)$, $s, t \in M$, as defined in eq.~\eqref{standDif2}, following derivations from \cite{Billard2005}, the asymptotic correlation of  $\hat{u}_{s}$ and $\hat{u}_{t}$, $\psi_{st} = \psi_{hi,jl} =\text{cor}(\hat{u}_{s},\hat{u}_{t})$, as $n\to \infty$,  is expressed by
\begin{equation}\label{corSep}
\psi_{st} = \psi_{hi,jl} =  (\omega_{hh | l}\,\omega_{jj | l})^{-1}[
(\omega_{hj | i} \,\omega_{il | j} + \omega_{hj | l}\, \omega_{il | h})
 + (\omega_{hl | i}\,\omega_{ij | l} + \omega_{hl | j}\,\omega_{ij | h})]/2,
\end{equation}
where $\omega_{hi | j} = r_{hi} - r_{hj}r_{ij}$ and $\omega_{hh | l} = 1 - r_{hl}^2$.

The difference of Fisher transformed coefficients also approximately follows a normal distribution
%\begin{equation}\label{deltaU}
%$\Delta \hat{u}_{ij}  \coloneqq (\hat{u}_{Y_{ij}} -\hat{u}_{X_{ij}}) \sim N(u_{Y_{ij}} - u_{X_{ij}}, 2(1-\psi_{ij})),\mbox{\hspace{0.1cm} for any \hspace{0.2cm}} i\neq j,$
$\Delta \hat{U}  \coloneqq (\hat{U}_{2} -\hat{U}_{1}) \sim N(U_{2} - U_{1}, \Psi_1 + \Psi_2 - 2\Psi_{12})$
%\end{equation}
where $\Psi_{12}$ describes the correlation between coefficients in $\hat{U}_{1}$ and $\hat{U}_{2}$.\ The diagonal elements $(\psi^{(12)}_{tt})$, $t\in M$,  are estimated by plugging-in the sample correlation coefficients in eq.\ \eqref{corSep}.\ This yields a consistent estimator of $(\psi^{(12)}_{tt})$ for large $n$ but produces non-negligible bias in the estimation for small $n$.\
Let $\hat{d}_{t}$ be the standardized expression of $\Delta \hat{u}_{t}$, such that
\begin{equation}\label{standDif}
\hat{d}_{t} = \Delta \hat{u}_{t}\{2(1 - \hat{\psi}^{(12)}_{tt})\}^{-1/2}, \quad t\in M, \quad \hat{D} = (\hat{d}_{t}).
\end{equation}
Under the null hypothesis of equality in the correlation matrices,  $\hat{d}_{t}$ has zero expected value and variance $(\sigma^2_{t})_n$ with  $( \sigma^2_t )_n \to 1$, $n\to \infty$ for any $t\in M$.\ Moreover, if $\psi^{(12)}_{tt}$ is known, then $\text{cov}( \hat d_t, \hat d_k)$ is proportional to $\psi^{(1)}_{tk} + \psi^{(2)}_{tk} - 2\psi^{(12)}_{tk}$, which is non-zero for some $k \neq t$, unless $R = I$.\

%From eq.\ \eqref{eq:ztransf} and eq.\ \eqref{standDif}, the  marginal distribution of $\hat{d}_{t}$, denoted by $\tilde{F}_{t}$,  depends on the distribution of  $g(\hat{r}_{t})$ and $\hat{\psi}_{t}$.\
%The Fisher transformation of sample correlation coefficients converges in distribution to a normal distribution.\
%Even for small sample sizes, it can produce a general good approximation (i.e. see first and second order statistics) but it is slightly inaccurate in representing the tail of the distribution.\ 
%Moreover, $[\hat{\psi}_{t}]$ are bounded between $(-1,1)$ but their distributions are not identical and difficult to approximate by a parametric family.\ Nevertheless, we have seen in simulated data that these tend to be short-tailed distributions.\ For instance, we  have observed that even for small $n$, the empirical distribution $\hat{F}_{t}$  is better approximated by a normal distribution than a student's t-distribution.\  
%Note that $\text{cor}( \hat d_t; \hat d_k) \neq 0$, for some $k \neq t$, unless $R = I$.\ For small sample sizes,  the coefficients $[\hat{\psi}_{t}]$ are also themselves correlated and generate more dependence in the matrix  $\hat{D}$.\ In Section 7 of the supplementary material we present a simulated data study to illustrate these issues in more detail.

\subsection{Test statistics}\label{testStatistics}
The three test statistics considered here are based on the elementwise standardized differences between transformed sample correlation coefficients in eq.~\eqref{standDif}.\ These are average of squares ($T_S$), extreme value ($T_M$) and sum of exceedances  ($T_E$) test statistics
%\begin{align}
%&T_{S} = m^{-1} \sum_{t \in M} \hat{d}_{t}^2, \label{eq:sumOfSquaresTest}\\
%&T_{M} = \max\limits_{t \in M} |\hat{d}_{t}|, \label{eq:maxTest}\\
%&T_{E}^{w}(u) = \sum_{t \in M} (|\hat{d}_{t}|-uw)^2  I( |\hat{d}_{t}| >u).\label{EX}
%\end{align}
\begin{equation}\label{TestStatistics}
T_{S} = m^{-1} \sum_{t \in M} \hat{d}_{t}^2, \,\,\,\,\,\,
T_{M} = \max\limits_{t \in M} |\hat{d}_{t}|, \,\,\,\,\,\,
T_{E}^{w}(u) = \sum_{t \in M} (|\hat{d}_{t}|-uw)^2  I( |\hat{d}_{t}| >u).
\end{equation}
In the sum of exceedances test, $w$ is either $0$ or $1$ and it is incorporated to weight the importance of high values over the threshold $u$.\

%We also consider as test statistic the maximum of the elements in $\hat{D}$ in absolute value:
%\label{eq:maxTest}
%The extreme value test ($T_{M}$) only needs for one correlation difference to be very large but generally when the correlation matrices are not equal there are several coefficients that change. Nevertheless, by using all the elements as for the average of squares test ($T_{S}$) we are likely to incorporate many spurious coefficients.
%Here we present a test statistic which is based on computing a partial sum of squares corresponding to exceedances above a high threshold $u$:
%\label{EX}
%where $w$ is either $0$ or $1$ and it is incorporated to weight the importance of high values over the threshold $u$.

\section{Null distributions and asymptotic power}\label{SEC3}

\subsection{Average of squares test}\label{aveSq}
%Let $\hat{D}$ be the $p\times p$ matrix with the  Fisher transform sample correlation coefficient differences with $(\hat{d}_t:t\in M)$ defined in eq.~\eqref{standDif}. 
The next lemma provides expressions for the expected value and variance of the average of squares test statistic $T_{S}$, which is defined in eq.~\eqref{TestStatistics}.

\begin{lem}[\emph{Expected value and variance of $T_S$}]
Let $\mu_2=\mathbf{E}(\hat{d}_{t}^2)$ and $\mu_4=\mathbf{E}(\hat{d}_{t}^4)$.\ Define $\bar{\gamma}_2 = 2(m^2 - m)^{-1}\sum_{t<h} \text{cov}(\hat{d}_{t}^2,\hat{d}_{h}^2)$.%, which measures the average of the lower triangular matrix elements of the covariance of $\hat{D}^2$.\ 
The expected value and variance of $T_{S}$ are expressed by
\begin{equation}\label{eq:varDep}
\mathbf{E}[T_{S}]= \mu_2; \quad  \text{var}(T_{S}) = (\mu_4 - \mu_2^2)/m + (1-1/m)\bar{\gamma}_2.
\end{equation}
\end{lem}
\begin{pro} 
The proof of lemma~\ref{eq:varDep} can be found in Section~\ref{proof1} of the appendix.\
\end{pro}
Under $H_0$, asymptotically with $n\to \infty$,  $\hat{d}_{t}^2 \sim \chi^2_1$, for any $t \in M$.\  Besides, for sufficiently large $n$, it follows from the properties of $\chi^2_1$ that $\mu_2 \doteq 1$ and $\mu_4 \doteq 3$.\
Let $\nu= \sum_{t<h} I[\text{cov}(\hat{d}_{t}^2,\hat{d}_{h}^2) \neq 0]$ be an integer ranging in $[0,m(m-1)/2]$.\ If $\text{cov}(\hat{d}_{t}^2,\hat{d}_{h}^2)\leq k$, for any $t<h$,  for a finite constant $k$, and $\nu/m \to 0$ as $m\to \infty$, then it follows that $\text{var}(T_{S}) = (2/m) (1 + O(\nu/m))$. 
% NAB: removed the following paragraph
%Under the asymptotic independence assumption, for large $m$, the leading terms are $\hat{\mathbf{E}}[T_{S}]=1$ and $\hat{\text{var}}(T_{S}) = 2/m$, and $T_{S}$ follows  a Gamma distribution with parameters $a=m/2$ (shape) and $b=2/m$ (scale).\ This is proved using the first and second order moments of a Gamma distributed random variable which ascertains that $\mathbf{E}(T_{S}) = ab$ and $\text{var}(T_{S}) =ab^2$.\
%%Then, the cumulative distribution function (CDF) of $T_{S}$ under $H_0$ is determined by $\Pr(T_{S} \leq x \mid H_0) \doteq \Gamma(x; a,b)$ where $\Gamma(\cdot; a, b)$ is the CDF of gamma distribution with parameters $a$ and $b$.

However, for a finite dimension, if the correlation matrices are not highly sparse, $\nu/m$  is not negligible and the dependence parameter $\bar{\gamma}_2$ must be incorporated to assure uniformity in the p-values of the test under $H_0$.\ Moreover, since an estimator for the covariance between Fisher transform sample correlations $\psi^{(12)}_{tt}$ (defined in eq.~\eqref{eq:ztransf}) is used, parameters $\mu_2$ and $\mu_4$ can differ slightly from their limiting values ($1$ and $3$) and should be estimated.\
For sufficiently large $m$ and $n$,   $T_{S}$ is well approximated by a normal distribution with parameters $\mu= \mu_2 $ and $\sigma^2 = (\mu_4 - \mu_2^2)/m + (1-1/m)\bar{\gamma}_2$ with $\Pr(T_{S} \leq x \mid H_0) \doteq \Phi(x; \mu, \sigma^2)$ where $\Phi(\cdot; \mu, \sigma^2)$ is the CDF of normal distribution with parameters $\mu$ and $\sigma^2$. Following the central limit theorem, the Gaussian approximation can be appropriate even when $n$ if parameters $\mu_2$ and $\mu_4$ are well specified (not approximated by their limiting values).

Hence, the null hypothesis is rejected at significance level $\alpha$ if the observed value of $T_S$ is greater than
\begin{equation}\label{eq:QuantAS}
t_{S,\alpha} \doteq \mu_2 + z_{\alpha}\sqrt{ (\mu_4 - \mu_2^2)/m + (1-1/m)\bar{\gamma}_2}.
\end{equation}
The following theorem shows a lower bound for the power of the average of squares test. 
\begin{thm}[\emph{Power of the average of squares test}]
Let $t_{S,\alpha}$ be asymptotic $\alpha$-quantile of the distribution for $T_{S}$ under $H_0$ defined by \eqref{eq:QuantAS} with $0<\alpha<1/2$.\  Under the alternative hypothesis, let
$\bar{\gamma}_2' = 2(m^2 - m)^{-1}\sum_{t<h} \text{cov}(\hat{d}_{t}^2,\hat{d}_{h}^2\mid H_1)$ and
$\delta_t = |g(r_{Y_t}) -g(r_{X_t})|$ with $\mathcal{S}_d = \{t\in M: \delta_t \neq 0\}$.\
% with $s=|S_d|$ being the number of coefficients in $S_d$ and $\rho_s = s/m$.
Denote  $\delta_0^2 = \sum_{t\in \mathcal{S}_d} \delta_t^2 $.\ If condition
\begin{equation}\label{condAS}
\delta_0^2 > z_{\alpha}\sqrt{2m} \{1 + (m-1)\bar{\gamma}_2/2)\}^{1/2}/(n-3)
\end{equation}
holds, then, as $n,m\to \infty$,  
$$
\Pr(T_{S} \geq t_{S,\alpha} \mid  H_1) \geq 1 -  \exp( -A^2/2)(1+o(1)).
$$
with
$$
A = \frac{ \frac{(n-3)}{m}\delta_0^2 - z_{\alpha} \sqrt{\frac{2}{m} \{1+ (m-1)\bar{\gamma}_2/2\}}}{(m^{-1/2}\{2+\frac{4s(n-3)}{m}\delta_0^2 +(m-1)\bar{\gamma}_2'\}^{1/2})} 
$$
\end{thm}
\begin{coro}\label{coro1}
For $\bar{\gamma}_2 < \nu\,k$ and $\nu/m = o(1)$, condition~\eqref{condAS} becomes $\delta_0^2 \gtrsim  \frac{m^{1/2}}{n}$ as $(n, m) \to \infty$.\ Under condition~\eqref{condAS}, when  $(n/\sqrt{m})\delta_0^2  \to \infty$, $\Pr(T_{S} \geq t_{S,\alpha} \mid  H_1) \to 1$.\
\end{coro}

%Using concentration inequality $P(T_{S} - E T_{S}\geq \lambda) \leq \exp\{ - \lambda^2/[]\} $

%The expressions for expected value and variance assuming independence between elements in $\hat{D}$  are then %given by $\mathbf{E}[T_{S}]= \mu_2$ and $\text{var}(T_{S}) = (\mu_4 - \mu_2^2)/m$ and the expressions for $a$ %and $b$ defined before hold. The p-value for the test statistic is calculated by
%\begin{equation}\label{asyPvalMax}
%\mbox{p-val} = \Pr(T_{S} \geq T_{obs}\mid H_0) = 1 - \Gamma(T_{obs}; a,b),
%\end{equation}
%where $\Gamma(\cdot; a, b)$ is the cumulative distribution of a gamma distribution with parameters $a$ and $b$. %Alternatively, we could approximate $T_{S}$ by a normal distribution with $\mu=1$ and $\sigma^2 = 2/m$ and p-value given by
%\begin{equation}\label{asyPvalMax2}
%\mbox{p-val} =  1- \Phi((T_{obs}-1)(2/m)^{-1/2}),
%\end{equation}
%where $\Phi$ is the cumulative distribution of a standardised normal distribution.

\subsection{Extreme value test}\label{maxTest}
In this section we provide a heuristic approach to
approximating the limiting distribution of $T_M$, defined in eq.~\eqref{TestStatistics}, based on two key
assumptions: $(i)$ we suppose that the sample size $n$ is sufficiently
large so that $(\hat{d}_t\,:\, t\in M)$ has a Gaussian distribution with
standard $N(0,1)$ margins and $(ii)$ we assume
\begin{IEEEeqnarray}{rCl}
  &&\max_{t<s\in M} \lvert \text{cov}(\hat{d}_{t},\hat{d}_{s})\lvert
  \,< \,1 \quad \text{and} \quad \nu_t = \sum_{s\in M \setminus t}
  I\{\text{cov}(\hat{d}_{t},\hat{d}_{s}) \neq 0\} = O(m^{\eta_t}),
  \label{eq:weak_dep}
\end{IEEEeqnarray}
for some $\eta_t\in(0,1)$, $t\in M$.\ Condition~\eqref{eq:weak_dep} implies
that no two elements of  $(\hat{d}_t)$ are perfectly dependent
and that there is sufficiently weak dependence structure in the
process.\ If condition~\eqref{eq:weak_dep}  holds, then adapted versions of extreme value limits for
non-stationary Gaussian processes apply \citep{Leadbetter1983},
i.e., there exist location and scale functions $\mu(m)\in \mathbb{R}$
and $\sigma(m)>0$, such that
\begin{IEEEeqnarray}{rCl}
  \lim_{m\rightarrow \infty}\Pr\left(\frac{T_{M}-\mu(m)}{\sigma(m)} <
    x\,\Big|\, H_0\right) %\nonumber
  %&=& \lim_{m\rightarrow \infty}\left[2\Phi\left\{x(m)\right\} -1\right]^{m}\nonumber\\
  &=& \exp\left\{-\exp\left(-x\right)\right\},    \label{depMax1}
%  \label{depMax1}
 \end{IEEEeqnarray}
describes a Gumbel distribution with $\mu(m)+\sigma(m)\,x \rightarrow \infty$, as
 $m\rightarrow \infty$, for all $x$.
% We do not give an exact proof of this statement but we note that
We note that a similar type of extreme value limits are
 obtained in \cite{Cai2013a} for the less general setting where
 $(Y_k^{(1)},Y_k^{(2)})$ in expression~\eqref{eq:jointCorHT} are independent.\
 Additionally, our empirical findings from simulations confirm that
 this is a reasonable approximation for the distribution of $T_M$
 provided $n$ and $m$ are sufficiently large.\ To back up this
 result, we illustrate in Appendix \ref{GumbelApprox} how
 condition~\eqref{eq:weak_dep} links with \cite{Leadbetter1983} conditions
 for convergence of the maximum of non-stationary Gaussian processes.\

% We consider $(\hat{d}_t:t\in M)$ as a stochastic
% process defined on $M=\{i,j \in \mathbb{N} \,:\, i < j\}$ of size $m$
% defines a non-stationary real valued stochastic process with common
% marginal distributions, which for sufficiently large sample size and
% for all $t\in M$ it is seen to be well approximated by a standard
% normal distribution\footnote{(which we denote by $Z$) \color{red}
%   somewhere else maybe more useful?\color{black}}.

%Under the stated assumptions,
In real applications, where $m$ is finite, limit expression~\eqref{depMax1}  may fail to approximate
the distribution of $T_M$  in two respects. Firstly, it is known that the rate
of convergence to the limit distribution is very slow.
%.\ In practical applications, this approximation
 %and bounded by
%\begin{IEEEeqnarray*}{rCl}
%\sup_{x\in \mathbb{R}}  \Bigg|\Pr\left(\frac{T_{M}-\mu(m)}{\sigma(m)} < x\,\Big|\, H_0\right)- \exp\left\{-\exp\left(-x\right)\right\}\Bigg| &\geq&
%  O\left(\frac{1
%    }{\log m}\right).
%\end{IEEEeqnarray*}
Secondly, its form is independent of the dependence structure of the
process $(\hat{d}_t\,:\,t\in M)$, a result that stems from the joint
tail properties of the multivariate Gaussian distribution \citep{Sibuya1960,  Tiago1962}.\

An improved approximation that does take into account the dependence characteristics can be obtained
from a sub-asymptotic correction \citep{Eastoe2012},
\begin{equation}
  \Pr\left(\frac{T_{M}-\mu(m)}{\sigma(m)} < x\,\Big|\, H_0\right)\doteq \exp\left\{-\left(\frac{m_E}{m}\right)\exp\left(-x\right)\right\},\quad \text{for large $m$},
  \label{depMaasx2}
\end{equation}
where $m_E=m_E(m,x)$ satisfies $m_E/m \rightarrow 1$, as
$m\rightarrow \infty$, for all $x\in \mathbb{R}$, and describes the
effective sample size of independent and identically distributed
$N(0,1)$ random variables whose maximum has the same distribution with
$T_M$.\ Note that the distribution of $T_{M}$ in eq.~\eqref{depMaasx2} is a Gumbel distribution as in eq.~\eqref{depMax1} but with an updated location parameter, say $\mu_{m_E}(m)$, which depends on $m_E$.\

Hence, the null hypothesis is rejected at significance level $\alpha$ if the observed value of $T_M$ is greater than
\begin{eqnarray}\label{eq:QuantM}
t_{M,\alpha} &\doteq& \mu_{m_E}(m) -  \sigma(m)\log(-\log(\alpha))\\
 &\sim & \sqrt{2\log(2m)} -[\log \theta_m + \log\{-\log(\alpha)\}]/\sqrt{2 \log(2m)}.\notag
\end{eqnarray}

The following theorem shows a lower bound for the power of the extreme value test 
\begin{thm}[\emph{Power of the extreme value test}]
Assume \eqref{eq:weak_dep} holds. Let $t_{M,\alpha}$ be the asymptotic $\alpha$-quantile of the distribution for $T_{M}$ under $H_0$ defined by \eqref{eq:QuantM} with $0<\alpha<1/2$.\ Under the alternative hypothesis, let  $\delta_t = |g(r_{Y_t}) -g(r_{X_t})|$ with $\mathcal{S}_d = \{t\in M: \delta_t \neq 0\}$.
If the following condition holds
\begin{equation}\label{condM}
\max_{t\in \mathcal{S}_d}\delta_t  >  \frac 1 {\sqrt{n-3}} \left[\sqrt{2\log(2m)} - \frac{\log\{-\log(\alpha)\}}{ \sqrt{2 \log(2m)}}\right],
\end{equation}
 then, as $n,m\to \infty$,
 \begin{align*}
\Pr(T_{M} \geq t_{M,\alpha} \mid  H_1)
&\geq  1 -    \exp\left[  -\frac{(n-3)}{2} \left\{  \max_{t \in \mathcal{S}_d} \delta_t -  \sqrt{\frac{2\log (2m)}{(n-3)}}\right\}^2\right](1+o(1)).
\end{align*}

If $s = |\mathcal{S}_d| \to \infty$ and
\begin{equation}\label{condMa}
\min_{t\in \mathcal{S}_d}\delta_t  >   \frac 1 {\sqrt{n-3}} \left[\sqrt{2\log(2m)} - \frac{\log\{-\log(\alpha)\}}{ \sqrt{2 \log(2m)}}\right],
\end{equation}
 then, as $n,m\to \infty$,
 \begin{align*}
\Pr(T_{M} \geq t_{M,\alpha} \mid  H_1) \geq 1 -  \exp\left\{ -e^{ - \sqrt{2(n-3)\log(2s)}\left[ \min_{t\in \mathcal{S}_d}\delta_t -\sqrt{\frac{2\log (2m)}{(n-3)}}\right] }\right\}(1+o(1)).
\end{align*}

\end{thm}

\begin{coro}
As $n, m \to \infty$, condition~\eqref{condM} becomes $\max_{t\in \mathcal{S}_d}\delta_t^2  \gtrsim   (2\log 2m)/(n-3)$.\ Under this condition,  if $\sqrt{n}(\max_{t\in \mathcal{S}_d}\delta_t - \sqrt{2\log(2m)/n}) \to \infty$,  $\Pr(T_{M} \geq t_{M,\alpha} \mid  H_1) \to 1$.

Similarly, as $n, m \to \infty$ and $s =|\mathcal{S}_d| \to \infty$, condition~\eqref{condMa} becomes $\min_{t\in \mathcal{S}_d}\delta_t^2  \gtrsim  (2\log 2m)/(n-3)$.\ Under this condition,  if $\sqrt{n\log s}(\min_{t\in \mathcal{S}_d}\delta_t - \sqrt{2\log(2m)/n}) \to \infty$,  $\Pr(T_{M} \geq t_{M,\alpha} \mid  H_1) \to 1$.
\end{coro}

\subsection{Sum of exceedances test}\label{excTest}
Let $\mathcal{S}_u = \{t\in M: |\hat{d}_t| \geq u\}$ be the set of exceedances above some  threshold $u\geq 0$, let $N_u = \text{Card}(\mathcal{S}_u)$ be the number of elements in $\mathcal{S}_u$ and recall that $m = p(p-1)/2$.  The cumulative distribution function of the test statistic $T_E$ under $H_0$ is
\begin{equation}\label{Te1}
\Pr(T_{E}^{w}(u)  < x \mid  H_0) = \sum_{k =1}^m \left[ \Pr(N_u = k \mid H_0)\,\, \Pr( T_{E}^{w}(u) < x \mid H_0, N_u = k)\right].
\end{equation}
We define several parameters that are used to determine the limiting distribution of $T_E$:
\begin{IEEEeqnarray}{rCl}
\gamma_{u_{tj}}^{(w)} &=& \text{cov}((|\hat{d}_t|-uw)^2,(|\hat{d}_j|-uw)^2 \mid  \hat{d}_t^2>u,\hat{d}_j^2>u, d_t = d_j =0), \nonumber\\
\eta_0 &=& \Pr( |\hat{d}_{t}| > u\mid d_t = 0),\label{sumEXCparam}\\
\phi_{tj} &=& \Pr(\hat{d}_t^2>u^2,\hat{d}_j^2>u^2\mid d_t =  d_j = 0), \,\,\, \bar{\phi} = [m(m-1)]^{-1}\sum_{t\neq j} \phi_{tj}. \nonumber
\end{IEEEeqnarray}
% $\eta_1 = \Pr( |\hat{d}_{t}|>u\mid H_1)$ and $m=p\,(p-1)/2$.\
%Below we consider two cases: asymptotic independence and dependence samples in $\hat{D}$. We also present the results for the two definitions of $T_{E}$ in eq. \eqref{EX}: $w=0$ and $w=1$.
%\
Let $\varphi$ and $\Phi$ be the density and cumulative distribution function of the standard normal distribution, respectively. For sufficiently large expected number of exceedances,  the central limit theorem yields  $\Pr (T_{E}^{w}(u) < x \mid  H_0)\doteq \Phi\{x, \mu(m,w), \sigma^2(m,w)\}$ for any $w=\{0,\,1\}$, with
\begin{equation}\label{musigma3}
\begin{cases}
                  \mu(m,w) = m\,\eta_0\,\mu_{w}\\
                 \sigma^2(m,w) = m\,\{ \eta_0\sigma_{w}^2 + \mu_{w}^2 (\eta_0-\bar{\phi}) \} + m^2\mu_{w}^2 (\bar{\phi} - \eta_0^2) + \sum_{t\neq j} \gamma_{u_{tj}}^{(w)} \phi_{tj},
\end{cases}
\end{equation}
where $\mu_{w}$ and $\sigma^2_{w}$ are defined for $w=0$ by
\begin{equation}\label{musigma1}
\begin{cases}
                  \mu_0        =  1+ u\, \varphi(u)/\{1-\Phi(u)\} \\
                 \sigma^2_0 =  3 + (u^3 + 3u)\,\varphi(u)/\{1-\Phi(u)\}  -\mu_0^2,
\end{cases}
\end{equation}
whereas for $w=1$ these are
\begin{equation}\label{musigma2}
\begin{cases}
                  \mu_1 = u^2+1 - u\, \varphi(u)/\{1-\Phi(u)\}  \\
                 \sigma^2_1 = 3+u^4+6u^2 - (5u+u^3)\,\varphi(u)/\{1-\Phi(u)\} -\mu_1^2.
\end{cases}
\end{equation}
[The derivation of equations~\eqref{musigma3}, \eqref{musigma1} and \eqref{musigma2} can be found in Section~\ref{proof3} of the Appendix].\
Note that if the elements in $\hat{D}$ are near independence, then $\bar{\phi} \approx \eta_0^2$, making the third term in the expression for the variance in eq.~\eqref{musigma3}  approximately zero, and the whole expression simplifies to $ \sigma^2(m,w) \doteq m\,\eta_0\{(1-\eta_0)\,\mu_{w}^2 + \sigma_{w}^2\}$.
% which determines the  distribution for the test statistic under asymptotic independence.\

The null hypothesis is rejected at significance level $\alpha$ if the observed value of $T_E^{(w)}$ is greater than
\begin{equation}\label{eq:QuantE}
t_{E,\alpha}^{(w)} \doteq \mu(m,w) + z_{\alpha}\sigma(m,w).
\end{equation}
The following theorem shows a lower bound for the power of the sum of exceedances test.
\begin{thm}[\emph{Power of the sum of exceedances test}]
Let $t_{E,\alpha}^{(w)}$ be the asymptotic $\alpha$-quantile of the distribution for $T_{E}^{(w)}$ under $H_0$ defined by \eqref{eq:QuantE} with $0<\alpha<1/2$ and $w$ being either $0$ or $1$.
Consider $\mu_0$ and $\mu_1$  defined by eq.~\eqref{musigma1} and eq.~\eqref{musigma2}, $\eta_0$ defined by eq.~\eqref{sumEXCparam} and $\sigma^2(m,w)$ defined by eq.~\eqref{musigma3}.\ Under the alternative hypothesis, let  $\delta_t = |g(r_{Y_t}) -g(r_{X_t})|$ with $\mathcal{S}_d = \{t\in M: \delta_t \neq 0\}$, $s=|\mathcal{S}_d|$,   $\eta_t =  \Pr( |\hat{d}_{t}| > u\mid d_t \neq 0)$ and  $\mu_{t_w} =  \mathbf{E}( (|\hat{d}_{t}| -wu)^2 \mid |\hat{d}_t|>u, d_t \neq 0)$. If the following condition holds
\begin{equation}\label{condE}
\sum_{t\in \mathcal{S}_d} \mu_{t_w}\eta_t  > s\eta_0\mu_w -z_\alpha \sigma(m,w),
\end{equation}
then the lower bound for the asymptotic power of sum of exceedances test, with $w=\{0, 1\}$, as $n,m\eta_0\to \infty$, is
\begin{align*}
&\Pr(T_{E}^{(w)} \geq t_{E,\alpha}^{(w)} \mid  H_1) \geq 1 -
  \exp(-B(\delta_t,s,u,n,m,w)^2/2)(1+o(1)),
\end{align*}
with 
\begin{equation}\label{powerFunction}
B(\delta_t,s,u,n,m,w) = \frac{\sum_{t\in \mathcal{S}_d}  \mu_{t_w} \eta_t - s\,\eta_0\,\mu_w -z_\alpha\,\sigma(m,w)}{
\sigma_{H_1}(m,w) },
\end{equation}
where  $\sigma^2_{H_1}(m,w)$ is defined in Section~\ref{asymExceP1} of the appendix.

\vspace{0.2cm}
\hspace{-0.55cm}Note: Gaussian approximation represents well the asymptotic power if and only if $m\eta_0$ is sufficiently large, with
$u<\sqrt{2\log 2m}$ being a necessary condition.
\end{thm}
\begin{coro}
Assume $\sigma^2(m,w) \doteq m\,\eta_0\{(1-\eta_0)\,\mu_{w}^2 + \sigma_{w}^2\}$. Let $u = u(\beta)$ with $\beta = 2(1-\Phi(u))$, and  let $\mathcal{S}_{du} = \{t\in M, |d_t|\gg u\}$ with $s_u = |\mathcal{S}_{du}|$. When  $(m, n, u) \to \infty$, under condition~\eqref{condE}, if $s_u = k\max(1, s\eta_0, (2m\eta_0)^{1/2})$ for some integer $k>0$, and $\delta_t^2\, (n/u^2) \to \infty$ for some $t\in \mathcal{S}_{du}$, $\Pr(T_{E}^{(w)} \geq t_{E,\alpha}^{(w)} \mid  H_1) \to 1$.

\begin{enumerate}
\item $u = 0$: recovery conditions coincide with the average of squares test (Section \ref{aveSq}).
\item $u = \sqrt{2\log 2m}-o(1)$: recovery conditions are similar to extreme value test (Section \ref{maxTest}).
\end{enumerate}
\end{coro}

\subsection{Threshold selection for sum of exceedances test}\label{thresholdSe}
\begin{figure}[h]
\begin{center}
 \begin{tabular}{cc}
     \subfloat[$w=0$]{\includegraphics[width=6.7cm, height = 5.5cm]{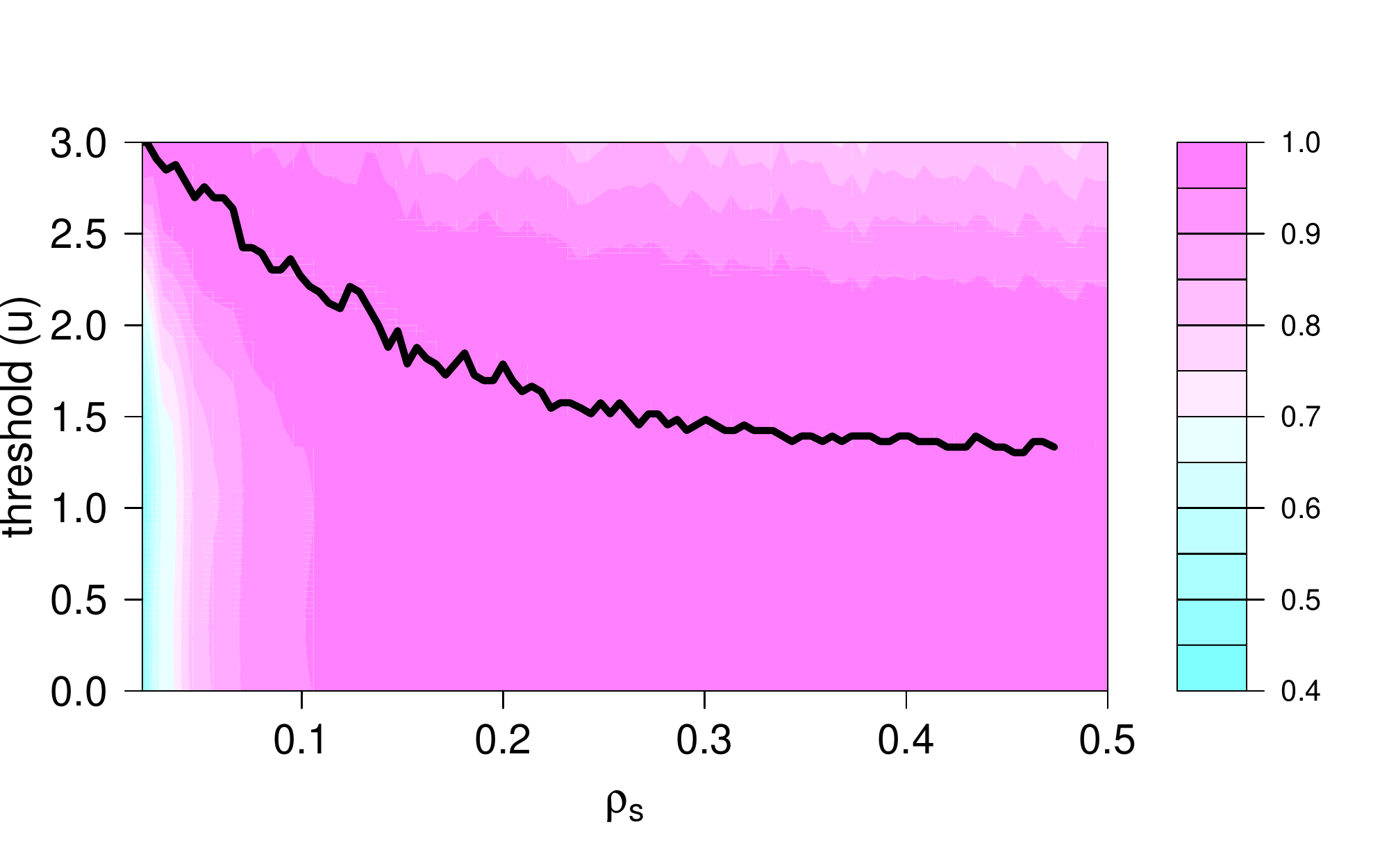}}&
         \subfloat[$w=1$]{\includegraphics[width=6.7cm, height = 5.5cm]{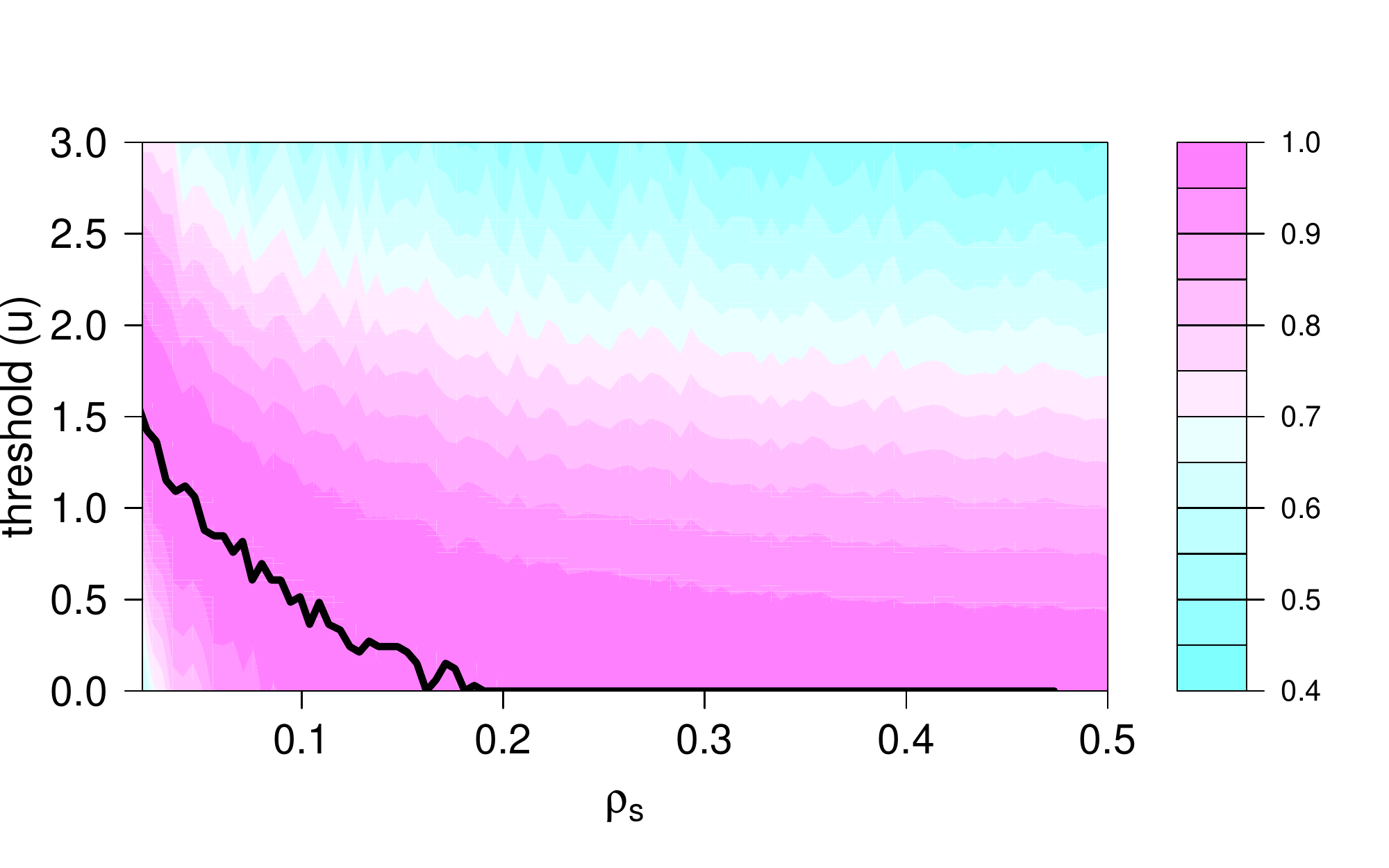}}\\
\end{tabular}
 \caption{ Relative power of sum of exceedances test with respect to threshold ($u$) and proportion of non-zero correlation differences ($\rho_s$) for (a) $w=0$ and (b) $w=1$.\ The black line corresponds to the threshold with highest power.\ }\label{Fig1}
\end{center}
\end{figure}

The threshold $u$ is key to find the test statistic that maximizes the power and its selection is the focus of attention of this section.\ Under notation in Theorem 3, $B(\delta_t,s,u,n,m,w)$ depends on parameters $n, m, w, u$ (known), and $s, \delta_t$ (unknown).\ Let $\rho_s = s/m$ be the  proportion of non-zero elements in $R_2-R_1$.\ To show the influence that $\rho_s$ has in the asymptotic power, the function $f$, defined in eq.~\eqref{powerFunction}, is evaluated for several values of $\rho_s$, $u$, with fixed sizes $n=100$, $m=10000$ and generating values of  $\delta_t$ from a Gamma$(a,b)$ distribution with parameters $a=3$ and $b=10$.\ In Figure \ref{Fig1}, the optimal threshold, defined by the value of $u$ that maximizes $B$, is decreasing with $\rho_s$ for both $w=0$ and $w=1$.\

\begin{figure}[h]
\begin{center}
 \begin{tabular}{ccc}
     \subfloat[$w=0$]{\includegraphics[width=4.4cm, height = 5.5cm]{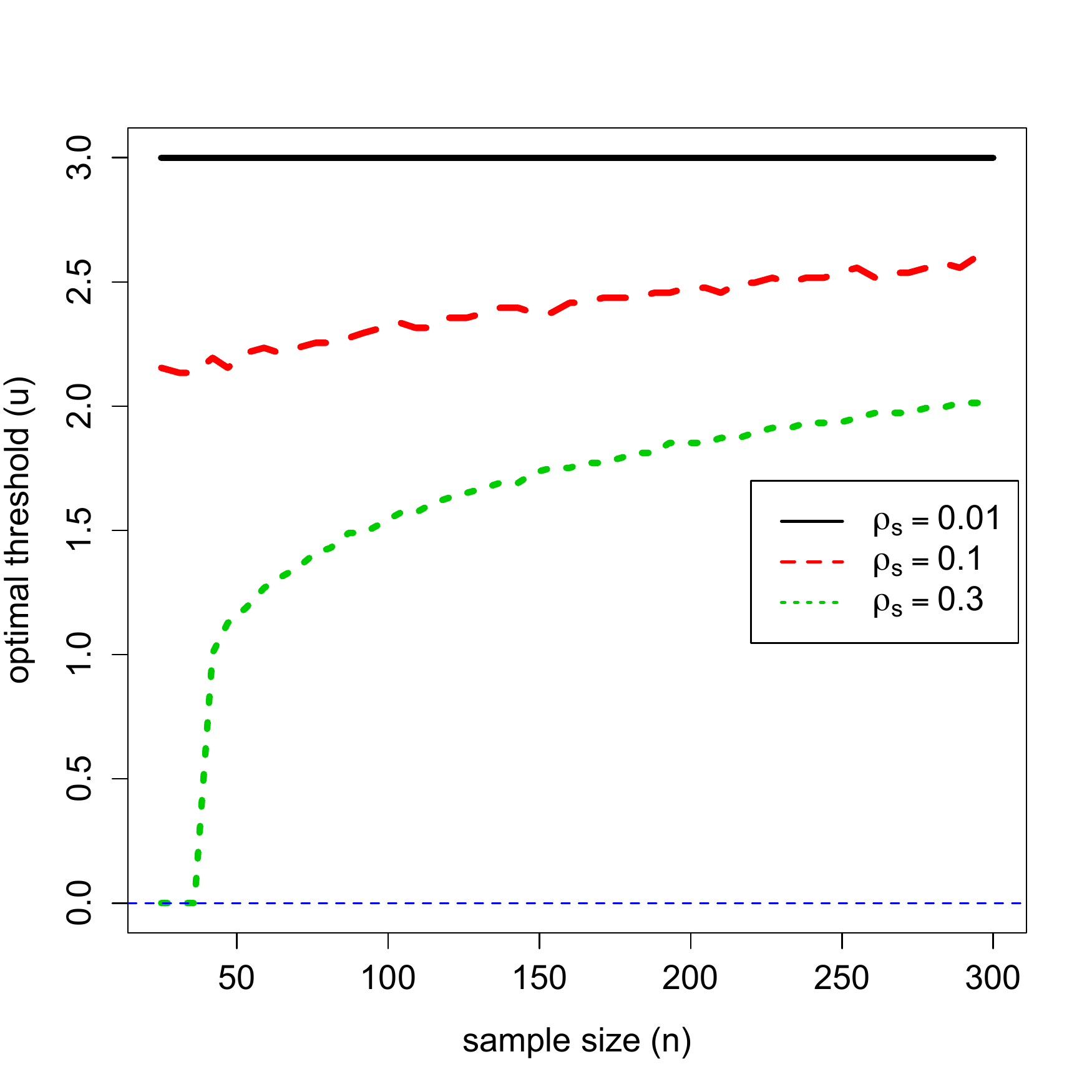}}&
     \subfloat[$w=1$]{\includegraphics[width=4.5cm, height = 5.5cm]{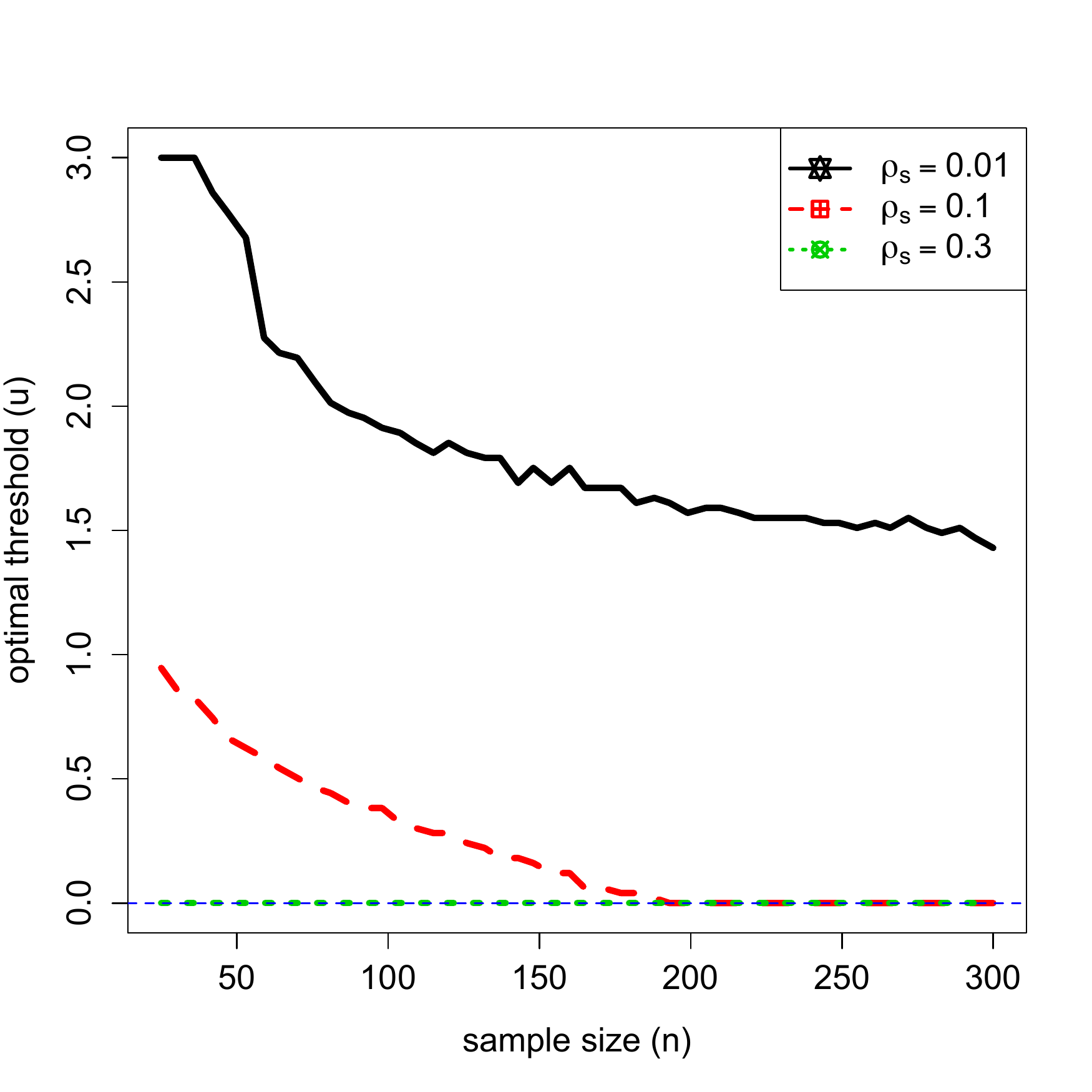}}&
     \subfloat[Difference]{\includegraphics[width=4.5cm, height = 5.5cm]{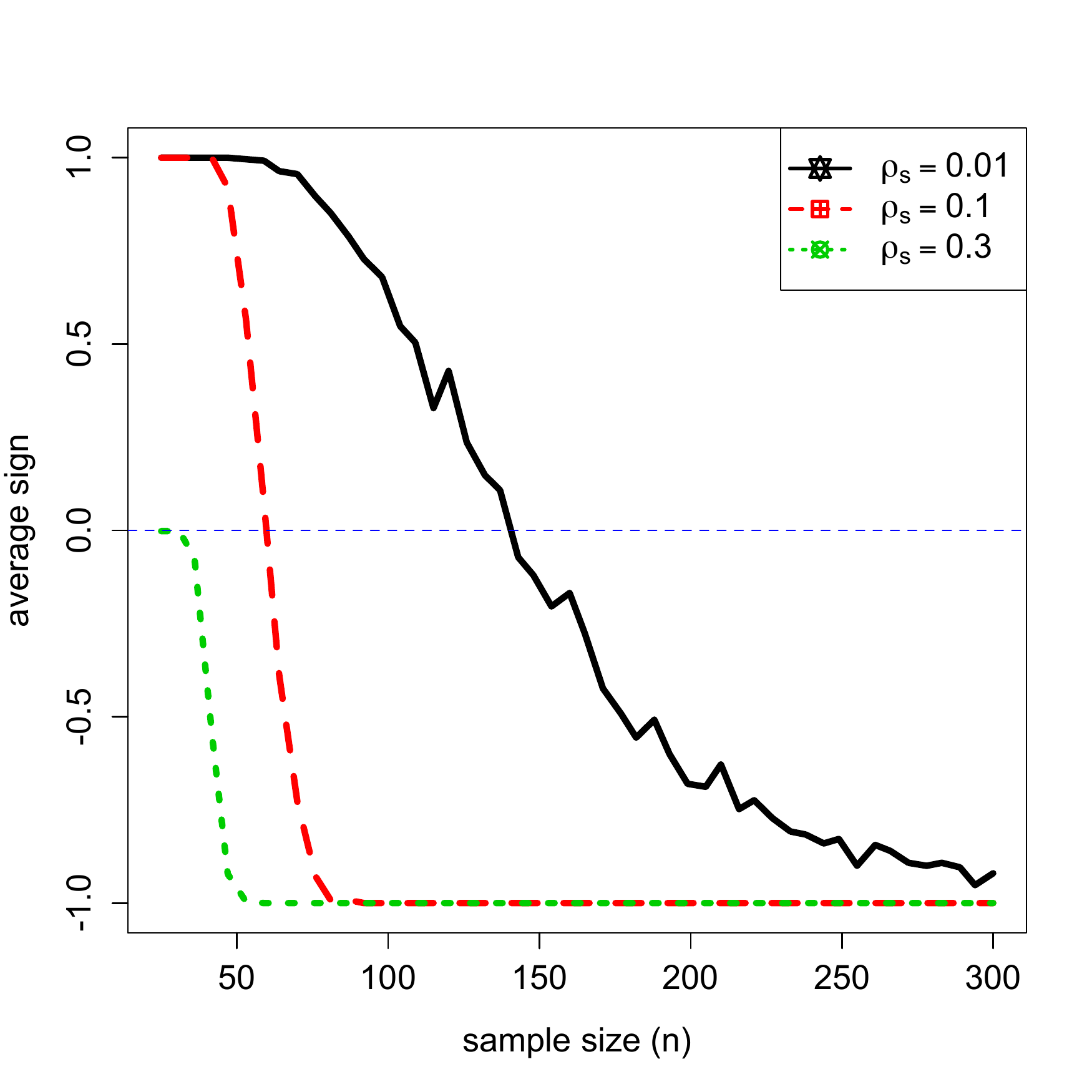}}
    \end{tabular}
 \caption{ Optimal threshold in sum of exceedances test  with respect to several values of the sample size for  (a) $w=0$ and (b) $w=1$.\ In (c) is shown the average sign for the difference between best power using $w=1$ and best power using $w=0$ over $1000$ simulated  sets of differential correlation coefficients. }\label{Fig2}
\end{center}
\end{figure}

Moreover, in panels (a) and (b) of Figure \ref{Fig2}, the optimal values for $u$ using a range of sample sizes and  three different values for $\rho_s \in \{0.01, 0.1, 0.3\}$ are obtained.\ We also considered several dimension sizes, but their impact on the threshold selection was very low and for simplicity we only show the cases for $m=1000$, which corresponds to $p \approx43-44$.\  For $w=0$, the optimal threshold increases with the sample size, whereas for $w=1$, the optimal threshold decreases with the sample size.\ In panel (c) of Figure \ref{Fig2}, we show the lower bound of the power differences between $w=0$ and $w=1$.\  We consider the best power for both $w=0$ and $w=1$ and then we take the difference between the two.\ In the figure we present the average sign of such power differences over $1000$ simulations for the set of parameters $(\delta_t: t \in \mathcal{S}_d)$.\ Only for small sample sizes ($n < 100$) and low $\rho_s$,  $w=1$ reaches better rates than $w=0$.\ Otherwise, $w=0$ dominates the asymptotic power.\

As Figure \ref{Fig1} and Figure \ref{Fig2} show, the fraction of zero elements in $R_2-R_1$ denoted by $\rho_s$ is essential to find the best threshold.\  We propose to find an estimator for $\rho_s$ using the q-values approach of \cite{Storey2015} where the input are approximated  p-values $2(1-\Phi(|\hat{d}_t|))$ for all $ t\in M$.\ Even though testing if $\rho_s = 0$ is the same as our hypothesis testing of $R_1 = R_2$, here we only use this testing procedure to find a first crude estimation of $\rho_s$.  This estimator is shown to be asymptotically unbiased with $n\to \infty$ but biased downwards when $\delta_t\sqrt{n-3}$ is small for all $t \in \mathcal{S}_d$ under mild dependence assumptions. However, in the application to biological data we generally have a  relatively small  $n$ and we have seen that  the dependence process in  $(\hat{d}_t\,:\, t\in M)$ can bias quite heavily the testing procedures in simulated data.% (see Section 3 from supplementary material).

The other unknown parameters are the Fisher transform correlation differences $\delta_t$, for all $t \in \mathcal{S}_d$.\ 
Below we propose a prior specification for $\delta_t$ to control the amount of elements that might be masked by the coefficients $\hat{d}_k,\, k \not\in \mathcal{S}_d$, when $\delta_k = 0$.\  However, other distributions or other specifications for the hyper-parameters could be employed instead.\
We  assume that $(\delta_t)$ are i.i.d.\ random variables with a known distribution, for instance we explore $\delta_t \sim \text{gamma}(a,b)$, with hyper-parameters satisfying $mode = (a-1)/b = Z_{\alpha}\,(n-3)^{-1/2}$, so the mode is assumed to be at the $1 - \alpha$ quantile of the marginal distribution of $\hat{d}_t\,(n-3)^{-1/2}$ under H$_0$.\ Moreover, we set the variance of the prior, $var = a/b^2$, so $a$ and $b$ are fully defined.\  
%For instance, in Figure \ref{FigHP} we show the density representation of some proposed distributions for $\delta_t$, as function of the selected variance, when $n=100$, and we compare them with the asymptotic distribution of  $|\hat{d}_t|\,(n-3)^{-1/2}$ for zero true elements.
% In Section \ref{excThreshold1} we carry out a sensitivity analysis to compare the performance of the test over different magnitudes for the variance.\

%\begin{figure}[h]
%\begin{center}
% \includegraphics[scale=0.37]{figures/hyperparameters.pdf}
% \caption{Some prior distributions for $\delta_t$ that are used to select the threshold for the sum of exceedances test. These are compared with the
% density distribution of $|\hat{d}_t|\,(n-3)^{-1/2}$ under $H_0$ for $n=100$ which is the black solid line.\ }\label{FigHP}
%\end{center}
%\end{figure}

We numerically integrate out $\delta_t$ from the function $B(\delta_t,s,u,n,m,w)$ defined in eq.~\eqref{powerFunction}  for threshold selection, i.e.,
$$
\hat{u}^w = \arg\max_u \int_{\Omega_{\delta_t}} B(\delta_t,m\hat{\rho}_s,u,n,m,w) p(\delta_t) \,d\delta_t.
$$
As final estimate we use the minimum between the optimal threshold and the $1-\alpha$ quantile of a standard normal distribution with default value $\alpha=0.05$ in order to prevent cases with infinite thresholds.\

\subsection{Estimation of dependence parameters and permutations based distributions}\label{perm}
Under $H_0$, $X_1,\ldots, X_n \sim N(0,R_X)$ and $Y_1, \ldots, Y_n \sim N(0,R_Y)$ with $R_X=R_Y$.\  In case $X_k$ and $Y_k$ were independent for all $k \in \{1,\ldots,n\}$, the elements in $[X_1,\ldots,X_n,Y_1,\ldots, Y_n]$ would be exchangeable (i.e., permutation invariant).\ For paired datasets,  $R_{XY}\neq 0$ and standard permutation methods are not suitable.\ Alternatively, we consider a resampling method which keeps
paired observations together: find $[(Z_1^{\pi_1},\ldots, Z_n^{\pi_n}),(Z_1^{\bar{\pi}_1},\ldots, Z_n^{\bar{\pi}_n})]$ where $\bar{\pi}_k=1-\pi_k$, and
$Z_k^{\pi_k} = X_i \mbox{ if } \pi_k = 0$ or $Z_k^{\pi_k} = Y_k \mbox{ if } \pi_k = 1$, with $\pi_k \sim \text{Bern}(1/2)$.\ The permutation process is repeated $B$ times and for each replicate ($i = 1, \ldots, B$)  the difference of Fisher transform correlation matrices,  defined in eq.~\eqref{standDif}, is calculated and denoted by $\hat{D}^{(i)}$.\ Finally, a $B\times m$ matrix $\tilde{D}$ is considered where row $i$ contains the lower triangular matrix of $\hat{D}^{(i)}$.\

%\vspace{0.5cm}
%\hspace{-0.8cm} {\it Average of squares test}\\
We denote $\tilde{D}^2$ by the elementwise product of the matrix $\tilde{D}$ and $\tilde{D}^4$ by the elementwise product of the matrix $\tilde{D}^2$. 
The parameters $\mu_2$, $\mu_4$ and $\bar{\gamma}_2$ for the average of squares test defined in eq.\ \eqref{eq:varDep} are estimated using permuted samples such that
$$
\hat{\mu}_2 = \frac{1}{mB}\sum_{i=1}^B \sum_{t=1}^m \tilde{D}_{it}^2, \,\,\, \hat{\mu}_4 =\frac{1}{mB}\sum_{i=1}^B \sum_{t=1}^m \tilde{D}_{it}^4,\,\,\,
$$
$$
\hat{\bar{\gamma}}_2 = \frac{2}{Bm(m-1)}\sum_{i=1}^B\sum_{t<h} \text{cov}(\tilde{D}_{it}^2, \tilde{D}_{ih}^2).
$$
%
%
%\vspace{0.5cm}
%\hspace{-0.8cm} {\it Extreme value test}\\
Regarding the extreme value test, for each replicate of the permutation process, $i=1,\ldots, B$, the maximum $\hat{T}_{M}^{(i)} =\max_{t \in M} |\tilde{D}_{it}|$ is computed so that for sufficiently large sample size $n$, $\hat{T}_{M}^{(i)}$  for all $i = 1, \ldots, B$ can be considered as an independent replicate of a Gumbel  distributed random variable with parameters $\mu_{m_E}(m)$ and $\sigma(m)$.\  The location parameter $\mu_{m_E}(m)$ of the Gumbel distribution is estimated by maximum likelihood.\ 
%The extremal index $\theta_m$, which is defined in eq.~\eqref{depMax2}, can also be estimated to obtain a measure of global dependency, for instance see the proposed methods in \cite{Ancona-Navarrete2000}.
%A non-parametric null distribution for $T_{M}$ is found by computing  $B$ test statistics $T_{M}^{(i)} = \max_{t \in M} |\tilde{D}_{it}|$, for $i=1,\ldots, B$, with $\Pr(T_{M} \leq x \mid H_0) \doteq B^{-1}\sum_{i=1}^B I(\hat{T}_{M}^{(i)} \leq x)$.\
%
%\vspace{0.5cm}
%\hspace{-0.8cm} {\it Exceedances-based test}\\
Besides, for the sum of exceedances test, the parameter $\sigma^2(m,w)$ defined in eq.~\eqref{musigma3} is estimated by maximum likelihood using permuted samples such that  $\Pr(T_{E}^{w}(u) < x \mid  H_0)  \doteq \Phi\{x,\mu(m,w),\hat{\sigma}^2(m,w)\}$ where the parameter $\mu(m,w)$ is also expressed in  eq.~\eqref{musigma3}.\  
%
%The non-parametric distribution is represented by  $B$ test statistics
% $\hat{T}_{E}^{(i)} =    \sum_{t \in \mathcal{S}_u} (\tilde{D}_{it} - uw)^2$, for $i=1,\ldots,B$, with
% $\Pr(T_{E} \leq s \mid H_0) \doteq B^{-1}\sum_{i=1}^B I(\hat{T}_{E}^{(i)} \leq s)$.\

A non-parametric null distribution for $T_{Q}$, $Q \in {S, M, E}$, based on permuted samples  is also considered by recording the value of  $B$ test statistics computed by $\hat{T}_{S}^{(i)} =   m^{-1} \sum_{t = 1}^m \tilde{D}_{it}^2$, $T_{M}^{(i)} = \max_{t \in M} |\tilde{D}_{it}|$ or $\hat{T}_{E}^{(i)} =    \sum_{t \in \mathcal{S}_u} (\tilde{D}_{it} - uw)^2$, for $i=1,\ldots,B$, with $\Pr(T_{Q} \leq x \mid H_0) \doteq B^{-1}\sum_{i=1}^B I(\hat{T}_{Q}^{(i)} \leq x)$.\

\subsection{Comparison of the tests}\label{testStatisticsTec}
Extreme value test is more powerful when it comes to sparse alternatives whereas sum of squares test is useful when the differential correlation matrix is non-sparse and the magnitude of the coefficients is small.\ The sum of exceedances test lies in between the other two tests.\ For threshold $u$ near zero, the test statistic is similar to the average of squares test and for $u\approx \sqrt{2\log m}$ it finds similar powers to the extreme value test.\ The weight $w$ is added to the expression of the sum of exceedances since the underlying test powers are complementary regarding sample sizes and number of  non-zero correlation  differences.\ For instance, for $w=1$ the test is  powerful for highly sparse differential correlation matrix  and small sample sizes (or small magnitude for the difference coefficients).\ Otherwise, $w=0$ achieves the most powerful test of the two.\ We consider a default value of $w=0$.  The theoretical results obtained in this section are completed empirically using simulated data  in Section \ref{SEC4}.

%%%%%%%%%%%%%%%%%%%%%%%%%%%%%%%%%%%%%%%%%%%%%%%%%%%%%%%%%%%%
%%%%%%%%%%%%%%%%%%%%%%%%%%%%%%%%%%%%%%%%%%%%%%%%%%%%%%%%%%%%
\section{Comparison of the tests for simulated data}\label{SEC4}
We analyze the accuracy of the proposed methods in simulated data sets.\ We study different structures for the correlation matrix $R$ directly (Section \ref{denseCor}) or indirectly by setting different graph structures for the precision matrix $\Omega = R^{-1}$ (Section \ref{powerLaw}).\

\subsection{Independent datasets, dense correlation matrices}\label{denseCor}
We can observe in real data, some groups of highly dependent genes whose underlying correlation matrix is non-sparse.\
In such a case, we argue that asymptotic independence tests are not reliable under H$_0$ even when the datasets are independent.\
We show this in simulated data by considering a dense correlation matrix denoted by $\tilde{R}$.\ This matrix is
obtained by the sample correlation matrix of a subset of  50 variables from the real dataset described in
Section \ref{SEC5}.\  In order to obtain a  positive definite matrix, we regularize $\tilde{R}$ by
\begin{equation}\label{Regu}
\Sigma = \tilde{R} + I\lambda,
\end{equation}
where $\lambda>0$.\ Note that as we increase $\lambda$,  off-diagonal elements of the correlation matrix decrease.\  

Data  $Y^{(1)}_k \sim N(0,\Sigma_1)$ and $Y^{(2)}_k \sim N(0,\Sigma_2)$, i.i.d.\ for all $k = 1,\ldots,n$ are generated using the following specifications for the covariance matrices: (i) under $H_0$, we consider $\Sigma_1 = \Sigma_2 = \Sigma$; (ii) under $H_1$, we consider  $\Sigma_1 = \Sigma$ and for $\Sigma_2$, we create a two-block diagonal matrix of sizes 40 and 10 by setting to zero the between-block covariance elements of the matrix $\Sigma$.  We refer to this model in the results presented  in Sections \ref{powerSize}   as model 1, which is applied for $n = 50, 100$ and $\lambda = 1/2, 1, 2, 3$.\

\subsection{Dependent datasets, sparse correlation matrices}\label{powerLaw}
Sparse correlation matrices are obtained by setting almost-block diagonal precision matrices, where each block  has a power-law underlying graph structure  \citep{Peng2009} and  some extra random connections between blocks. Let $A$ be the adjacency matrix with the non-zeros of the precision matrix, the coefficients of the precision matrix are simulated by
\begin{equation}
\Omega^{(0)}= [\omega_{ij}^{(0)}], \mbox{\hspace{0.5cm}} \omega_{ij}^{(0)} = \left\{ \begin{array}{r l l}
      & \text{Unif}(0.5,0.9) & \mbox{if $A_{ij} = 1$ with probability $ 0.5$ } ;\\
      &  \text{Unif}(-0.5,-0.9) & \mbox{if $A_{ij} = 1$ with probability $ 0.5$ }; \\
      &   0 & \mbox{if $A_{ij} = 0$}.
         \end{array} \right.
\label{eqSM11}
\end{equation}
Data  $(Y^{(1)}_k ,Y^{(2)}_k)\sim N(0,\Omega^{-1})$, i.i.d.\ for all $k = 1,\ldots,n$ are generated using a direct effect model \citep{Wit2015} with the following specifications for the joint precision matrix $\Omega$: (i) under $H_0$, $\Omega$ is determined by $\Omega_1 = \Omega^{(0)}$, $\Omega_2= \Omega^{(0)}$ and $\Omega_{12}$ being a diagonal matrix with  $(\Omega_{12})_{ii}  = 0.6$ for $\lfloor p/2\rfloor$ diagonal elements and $(\Omega_{12})_{ii} = 0$ for the other $\lceil p/2\rceil$; under $H_1$,  let  $D_1$ and $D_2$ be two different precision matrices which are generated with the same model as for $\Omega^{(0)}$. We consider $\Omega_1 = \text{diag}(\Omega^{(0)}, D_1, I)$, $\Omega_2 = \text{diag}(\Omega^{(0)}, I, D_2)$ and the same specification for $\Omega_{12}$ given under $H_0$.\  In both setting, to obtain a positive definite matrix,  we regularize
$\Omega$ by $\Omega = \Omega +  \lambda I$, with $\lambda$ such that the condition number of
 $\Omega$ is less than the number of nodes  \citep{Cai2011}.
We use $p =  70, 120, 210$ and sample sizes $n =  25, 50, 100, 200$.\ We refer to this model in the results presented in Section \ref{powerSize} as model 2.

\subsection{Power and size of the tests}\label{powerSize}
We use the average of squares test -S-, the extreme value test -M- and the sum of exceedances test -E- for both $w=0$ and $w=1$ (see definition in eq. \eqref{TestStatistics}) with threshold selected as defined in Section \ref{excTest}.\
We compute the empirical power of the tests defined as $\Pr(\mbox{Reject}\,\, H_0\mid  H_1 \mbox{ true})$ as well as the test size described by $\Pr(\mbox{Reject}\,\, H_0\mid  H_0 \mbox{ true})$ using significance level of $\alpha =0.05$.\
We approximate the null distributions by assuming linear independence between elements in $\hat{D}$ (denoted by AI) since it is computationally very fast.\ Moreover, we approximate the distributions estimating the dependence parameters using permuted samples
(AD) and also using a non-parametric distribution (NP) as described in Section \ref{perm}.\
For $w=1$ we only show the power of the non-parametric null distribution which  is labeled by E(NP)$^{(1)}$.\ Nevertheless, test sizes when  $w=1$ are seen to be similar to the ones provided when $w=0$.\

In Table \ref{tabPOWER3} we present the empirical approximations of power and size for the dense correlation matrices scenario (model 1).\  Generally, tests show a good trade off between false rejection and true rejection rates.\  For low regularization $\lambda$, as defined in~\eqref{Regu}, asymptotic linear independence tests are not suitable with empirical sizes being larger than the expected $0.05$.\ The average of squares test is the one that dominates the powers in this model for $\lambda\geq 2$ and gives similar results to the sum of exceedances test (with $w=0$) for $\lambda< 2$.\ Sum of exceedances test with $w=1$ achieves worse powers than the test with $w=0$ for large $\lambda$.

\begin{table}[t]
\caption{\label{tabPOWER3}Size, uniformity  and power of the test using model 1 -dense correlation matrices- ($\times 10^3$). Test statistics S (average of squares), M (extreme values) and E (exceedances with $w=0$ or $w=1$), and null distributions AI (asymptotic independence), AD (asymptotic dependence)  and NP (non-parametric) are compared at $\alpha = 0.05$ level.} % title of Table

\tiny
\centering
\begin{tabular}[h]{ l r r r r  r  r r r r}
&\multicolumn{4}{c}{n=50}&&\multicolumn{4}{c}{n=100}\\
\cline{2-5}\cline{7-10}
$\lambda$ & $0.5$ &$1$& $2$ & $3$ & &$0.5$ &$1$& $2$ & $3$\\
\hline
\multicolumn{10}{c}{Empirical size} \\
S(AD)&\textcolor{red}{62}& 50& 58& 53&& 52& 59& 60& 52\\
S(NP)&\textcolor{red}{61}& 47& 54& 52&& 53& 54& 57& 50\\
S(AI) &\textcolor{red}{306}& \textcolor{red}{238}& \textcolor{red}{192}& \textcolor{red}{133}&& \textcolor{red}{304}& \textcolor{red}{254}& \textcolor{red}{192}& \textcolor{red}{126}\\
M(AD) &45& 43& 49& 61&& 42& 48& 54& 50\\
M(NP)&51& 44& 47& 59&& 50& 50& 51& 48\\
M(AI) &\textcolor{red}{68}& 58& 59& \textcolor{red}{66}&& \textcolor{red}{62}& 54& 59& \textcolor{red}{61}\\
E(AD)$^{(0)}$&49&54& 59& 48&& 52& 50& 48& 54\\
E(NP)$^{(0)}$&54& 50& 60& 55&& 46& 60& 46& 58\\
E(AI)$^{(0)}$ &\textcolor{red}{103}& \textcolor{red}{126}& \textcolor{red}{92}& \textcolor{red}{86}&& \textcolor{red}{200}& \textcolor{red}{158}& \textcolor{red}{121}& \textcolor{red}{88}\\

\multicolumn{10}{c}{ks.test p-value to test for uniformity in the correlation test p-values} \\
S(AD)&247 &23 &716 &317 &&72 &400 &151 &79\\
S(NP)&432 &15 &134 &181 &&62 &432 &500 &148\\
S(AI)&\textcolor{red}{0} &\textcolor{red}{0} &\textcolor{red}{0} &\textcolor{red}{0} &&\textcolor{red}{0} &\textcolor{red}{0} &\textcolor{red}{0} &\textcolor{red}{0}\\
M(AD) &865 &121 &835 &426 &&147 &52 &245 &646\\
M(NP)&936 &69 &400 &969 &&181 &48 &288 &181\\
M(AI)&\textcolor{red}{0} &\textcolor{red}{0} &24 &27 &&\textcolor{red}{0} &\textcolor{red}{0} &193 &150\\
E(AD)$^{(0)}$&51 &416 &779 &211 &&231 &123 & 532 &883 \\
E(NP)$^{(0)}$&288 &618 &241 &500 &&400 &723 &648 &785\\
E(AI)$^{(0)}$&\textcolor{red}{0} &\textcolor{red}{0} &\textcolor{red}{0} &\textcolor{red}{0} &&\textcolor{red}{0} &\textcolor{red}{0} &\textcolor{red}{0} &\textcolor{red}{0}\\

\multicolumn{10}{c}{Empirical power} \\
S(AD) & 890 &690 &342 &240&&\textbf{998}& \textbf{992} &802& 542\\
S(NP)& 897 &684 &\textbf{380} &\textbf{250}&&\textbf{998}& \textbf{992}& \textbf{806} &\textbf{574}\\
%S(AI) & 908 &588& 306 &236 && 998 &990 &802 &450\\

M(AD)&667& 270& 110 &109&&996 &758 &250 &122 \\
M(NP)&652& 280 &106 &105&&996 &766& 254& 118\\
%M(AI) & 702& 310 &126 &072&&996 &796& 248& 130\\

E(AI)$^{(0)}$ &\textbf{950} &735& 374 &202&&\textbf{998} &\textbf{992} &790 &447\\
E(NP)$^{(0)}$&943& \textbf{723}& \textbf{380}& 223 && \textbf{998}& \textbf{992}& 787& 442\\
%E(AI)$^{(0)}$ &973& 692 &251 &126 &&998 &972 &676 &346\\
E(NP)$^{(1)}$&940& 692 &304 &143 &&\textbf{998} &\textbf{992} &687 &413\\
\multicolumn{10}{c}{Estimated $\theta$} \\
$\hat{\theta}_{m}$ &.593 &.843 &.915 &.955&&.574 &.828 &.912 &.953
\end{tabular}
\end{table}

In Table \ref{tabPOWER2} we show a similar analysis for dependent datasets with sparse correlation matrices.\ Null distributions accounting for dependence (AD and NP) achieve better estimates of the size than asymptotic linear independence tests.\ Particularly, in the average of squares and sum of exceedances tests adjusting for dependence is desired to obtain a good representation of the null distribution.\ The asymptotic linear independence extreme value test finds good estimates for the size.\ It is slightly conservative for large p-values but these do not affect the evidence interpretation.\ Hence, for sparse dependence structures, the asymptotic extreme value test could be used to speed up the process.\ The sum of exceedances test with $w=1$ (i.e., see NP) produces consistently the highest powers among the three tests.\ Contrarily of what we observe in Table  \ref{tabPOWER3}, the test with $w=1$ gives better results than the one with $w=0$.\ Moreover, the extreme value test provides higher powers than the average of squares for large sample sizes.

\begin{table}[t]
\caption{Size, uniformity  and power of the test using model 2 -sparse correlation matrices- ($\times 10^3$).\ Test statistics S (average of squares), M (extreme value) and E (sum of exceedances with $w=0$ or $w=1$), and null distributions AI (asymptotic independence), AD (asymptotic dependence) and NP (non-parametric) are compared at $\alpha = 0.05$ level. }\label{tabPOWER2}

\tiny
\centering
\begin{tabular}{ l r r r r  r  r r r r r rrrr}
&\multicolumn{4}{c}{p=70}&&\multicolumn{4}{c}{p=120}&&\multicolumn{4}{c}{p=210}\\
\cline{2-5}\cline{7-10}\cline{12-15}
n & $50$ &$100$& $200$ & $500$ & &$50$ &$100$& $200$ & $500$& &$50$ &$100$& $200$ & $500$\\
\hline
\multicolumn{15}{c}{Empirical size} \\
S(AD) &50& 50& 50& 52 && 49& 42& 56& 52&&\textcolor{red}{38}& 46& 48& 54\\
S(NP)&58 &54 &50 &52 &&55& 48& 58& 50&&
52 &50 &50 &53\\
S(AI)&\textcolor{red}{32}& 58& \textcolor{red}{78}& \textcolor{red}{78}&&\textcolor{red}{22}& 40& \textcolor{red}{62}& \textcolor{red}{69}&&\textcolor{red}{4}& \textcolor{red}{26}& 44& \textcolor{red}{62}\\
M(AD)&55 &46 &51 &58 &&48& 54& 46& 48&&
48 &50 &56 &44\\
M(NP)&55 &44 &51 &57 &&48& 54& 46& 47&&
47 &51 &54 &44\\
M(AI) &60 &41 &47 &54 &&56& 57& 47& 47&&
\textcolor{red}{62} &54 &58 &42\\
E(AD)$^{(0)}$&50 &50 &52 &51 &&56& 54& 56& 44&&50 &43 &46 &53\\
E(NP)$^{(0)}$&47 &48 &49 &49 && 52& 53& 54& 44&& 48 &46 &47 &52\\
E(AI)$^{(0)}$ &56 &42 &\textcolor{red}{38} &\textcolor{red}{66} &&\textcolor{red}{66}& 47& 46& 52&&\textcolor{red}{64} &52 &54 &46\\

\multicolumn{15}{c}{ks.test p-value to test for uniformity in the correlation test p-values} \\
S(AD) &\textcolor{red}{1}& 376& 37& 895&& \textcolor{red}{0}& 929& 351& 31&& \textcolor{red}{0}& \textcolor{red}{0}& 886& 286\\
S(NP) &\textcolor{red}{5}& 536& 29& 794&& \textcolor{red}{0}& 648& 370& 48&& \textcolor{red}{0}& \textcolor{red}{0}& 500& 164\\
S(AI)  &\textcolor{red}{0}& \textcolor{red}{0}& \textcolor{red}{0}& \textcolor{red}{0}&& \textcolor{red}{0}& \textcolor{red}{0}& \textcolor{red}{0}& \textcolor{red}{1}&& \textcolor{red}{0}& \textcolor{red}{0}& \textcolor{red}{0}& \textcolor{red}{0} \\
M(AD) &58& 662& 528& 266&& 701& 836& 917& 423&& \textcolor{red}{5}& 837& 50& 498 \\
M(NP)&87& 500& 466& 341&& 648& 859& 936 &241&& \textcolor{red}{3}& 723& 33& 341\\
M(AI) &173& 255& \textcolor{red}{19}& 798&& 513& 241& 298& 78&& 435& 701& \textcolor{red}{19}& 267 \\
E(AD)$^{(0)}$ &888& 58& 914& 374&& 155& 819& 725& 349&& 598& 191& 85& 42 \\
E(NP)$^{(0)}$&43& 536& 913& 263&& 43& 648& 794& 466&& 610& 988& 241& 466 \\
E(AI)$^{(0)}$ &138& 360& \textcolor{red}{10}& 135&& 28& 207& 856& 39&& \textcolor{red}{0}&\textcolor{red}{ 5}& 100 &42 \\

\multicolumn{15}{c}{Empirical power}\\% for model 2.1: cluster difference network pattern} \\
S(AD)&60 &144& 437 &730 &&78& 88 &178 &398 && 4 &78 &152& 439\\
S(NP)&62 &150& 430& 720&&96 &\textbf{106}& 182& 404&&\textbf{86} &\textbf{94}& 160&440\\
%S(AI)&  68 &152 &331& 630&&87 &111& 162 &401&& 82 &91 &154&391\\

M(AD) &76& 220 &715 &944&&68 &76 &176 &722 && 42 &72& 180&651\\
M(NP) &82 &228&706& 950&&60 &58& 174& 710 && 44 &72 &174&649\\
%M(AI) &64 &180 &458 &894&&76& 76 &180 &714 && 56& 74 &168&632\\

E(AD)$^{(0)}$ &101&200 &631 &910&&80& 82 &170 &520 &&70 &74 &180&550\\
E(AI)$^{(0)}$ &\textbf{102} &204& 615& 960&&82& 80 &180& 544 &&72 &76 &182&534\\
%E(AI)$^{(0)}$  & 78 &261 &598& 954&&106 &93& 265 &819 &&72& 83 &214&801\\
E(NP)$^{(1)}$ &94& \textbf{316}& \textbf{800}& \textbf{984}&&\textbf{102} &94& \textbf{272} &\textbf{816} &&70 &84 &\textbf{232}&\textbf{836}\\

%\multicolumn{15}{c}{Empirical power for model 2.2: random difference network pattern} \\
%S(AD)& 90& 218& 424& 782&&116& 184 &524 &812&&  98 &196 &406& 874\\
%S(NP)& \textbf{112}& 234& 426& 778&&136& 186& 528 &814&&152 &210& 418 &876\\
%S(AI) & 102& 231& 436& 758&&112& 184& 508 &824&&122 &192 &405 &866\\

%M(AD) & 80 &200 &404 &888&&56 &178 &614 &928&&92 &230 &592 &974\\
%M(NP)&78&186 &404 &886&&56 &174 &614 &930&&88 &218 &594 &972\\
%M(AI) & 64& 76 &398 &886&&46 &166 &596 &928&& 98 &226 &578 &974\\
%
%E(AD)$^{(0)}$ &88 &262 &544& 922&&92& 240 &756 &\textbf{968}&&122& 314 &716 &990\\
%E(NP)$^{(0)}$ &86&\textbf{ 272} &\textbf{552}& \textbf{924}&&\textbf{118} &\textbf{252}&\textbf{ 776} %&954&&\textbf{166}&\textbf{324}& \textbf{728}& \textbf{994}\\
%E(AI)$^{(0)}$  & 76 &242 &572& 914&&108& 239 &716 &964&&154 &313 &705 &993\\
%E(NP)$^{(1)}$ &98& 238&448& 822&&126 &206& 608& 884 &&136 &234 &468 &910\\%

\multicolumn{15}{c}{Estimated $\theta$} \\
$\hat{\theta}_{m}$ &.790& .871& .908& .943&&.788& .848& .913& .945 &&.770& .841& .903 &.937
\end{tabular}
\end{table}

We also analyze the behavior of the tests with respect to the proportion of non-zero correlation differences $\rho_s$.\  In a global analysis, we compute the average power for small proportions ($\rho_s \leq 0.3$) and large proportions ($\rho_s>0.3$) using the three test statistics.\ The sum of exceedances test has average powers $0.426$ and $0.543$ respectively, the extreme value test obtains $0.373$ and $0.465$, and the average of squares produces $0.312$ and $0.477$.\ It is $T_{S}$ that benefits the most from the increase of difference coefficients.

For model 1 (dense difference correlations matrix), the correlation between p-values for the same test statistic using both non-parametric and asymptotic null distributions is very high (around $0.994$ in average) whereas the average correlation between extreme value and average of squares p-values is  $ [0.61, 0.48, 0.36, 0.30] $ in the four regularization parameters used.\ The p-values for the sum of exceedances test (for both $w$), seem to be more correlated to the p-values for the other two tests with $[0.91, 0.88, 0.82, 0.75] $ against the average of squares and $[0.75, 0.63, 0.55, 0.52]$ against the extreme value.\ For model 2 (sparse difference correlation matrix), the correlations are smaller with an average of $[0.19, 0.12, 0.07]$ between average of squares and extreme value p-values for the three dimensions used, $[0.55, 0.39, 0.27]$ between average of squares and exceedances and $[0.49, 0.49, 0.48]$ between extreme value and exceedances.

We estimate the extremal index $\theta_m$, which quantifies the dependence structure over high exceedances, and it is defined in Section \ref{maxTest}.\ In the sparse model 2, the average estimated $\theta_m$ gets close to 1 as the sample size increases.\ For large $n$, we could assume that $\theta_m$ is equal to 1 and use the asymptotic approximation which would speed up the results.\ However, for dense correlations as in model 1, $\theta_m$ can be quite small ($\approx 0.6$ for small regularization $\lambda$) and permutations-based tests should be used instead.
%\newpage

%\vspace{-0.6cm}

%\begin{table}[h]
%\scriptsize
%\begin{center}
%\caption{Efficiency of the test defined as the explained power between estimated threshold and best threshold. } % title of Table
%\begin{tabular}{ l r r r r r rrrr r rrrr }
%&\multicolumn{4}{c}{n=25}&&\multicolumn{4}{c}{n=50}&&\multicolumn{4}{c}{n=100}\\
%\cline{2-5}\cline{7-10}\cline{12-15}
%$\rho_s$ & $0.01$ &$0.08$& $0.15$ & $0.23$ & &$0.01$ &$0.08$& $0.15$ & $0.23$ && $0.01$ &$0.08$& $0.15$ & $0.23$\\
%\hline
%$\delta/2$ & 99 & 100 &  100 &100 && 100&100&100&100&&100&100&100&100\\
%$\delta$ &100 &100 &100 &100 &&100&100&100&100&&100&100&100&100\\
%$3\delta/2$ & 98&92&94 & 99&&100&100&100&100&&100&100&100&100\\
%$2\delta$ & 94& 86& 79 &85&&98 &97&99&100&&100&100&100&100
%\end{tabular}
%\label{tabPOWERfa3}
%\end{center}
%\end{table}

%%%%%%%%%%%%%%%%%%%%%%%%%%%%%%%%%%%%%%%%%%%%%%%%%%%%%%%%%%%%
%%%%%%%%%%%%%%%%%%%%%%%%%%%%%%%%%%%%%%%%%%%%%%%%%%%%%%%%%%%%

\section{Application to colon cancer gene expression data}\label{SEC5}
We apply the methods to a  case study of gene expression data  which can be downloaded at \url{http://www.ebi.ac.uk/arrayexpress/} and that
it is presented in \cite{Hinoue2012}.\ A total of 25 patients are examined, the gene expression profiling is obtained in each one of them for a colorectal tumor sample and its healthy adjacent colonic tissue: in total 50 samples and  $24,526$ genes.\

We use the equality between correlation matrices tests for multiples subgroup of genes (of the $\num{25e3}$).\ We are particularly interested in knowing how standard gene pathways change in different medical conditions.\ To assess which biological processes might be linked to changes in the gene connections we download $1,320$ gene sets from the MSig database (Subramanian et al., 2005), which represent canonical pathways compiled from two
 sources: KeGG (Kanehisa et al., 2016) and Reactome (Milacic et al., 2012).\ Then we compare correlations in the two medical conditions by only considering genes in each of the pathways.\ Hence, we test 1,320 different correlation matrices.\ Note that in the original data some genes are represented by more than one probe/sequence (these are not identical, so they are not merely technical replicates), in order to compare correlation matrices, we take the average of these probe/sequence for the same gene.\

\begin{figure}[h]
\begin{center}
 \begin{tabular}{cc}
    \subfloat[p-values]{\includegraphics[scale=0.38]{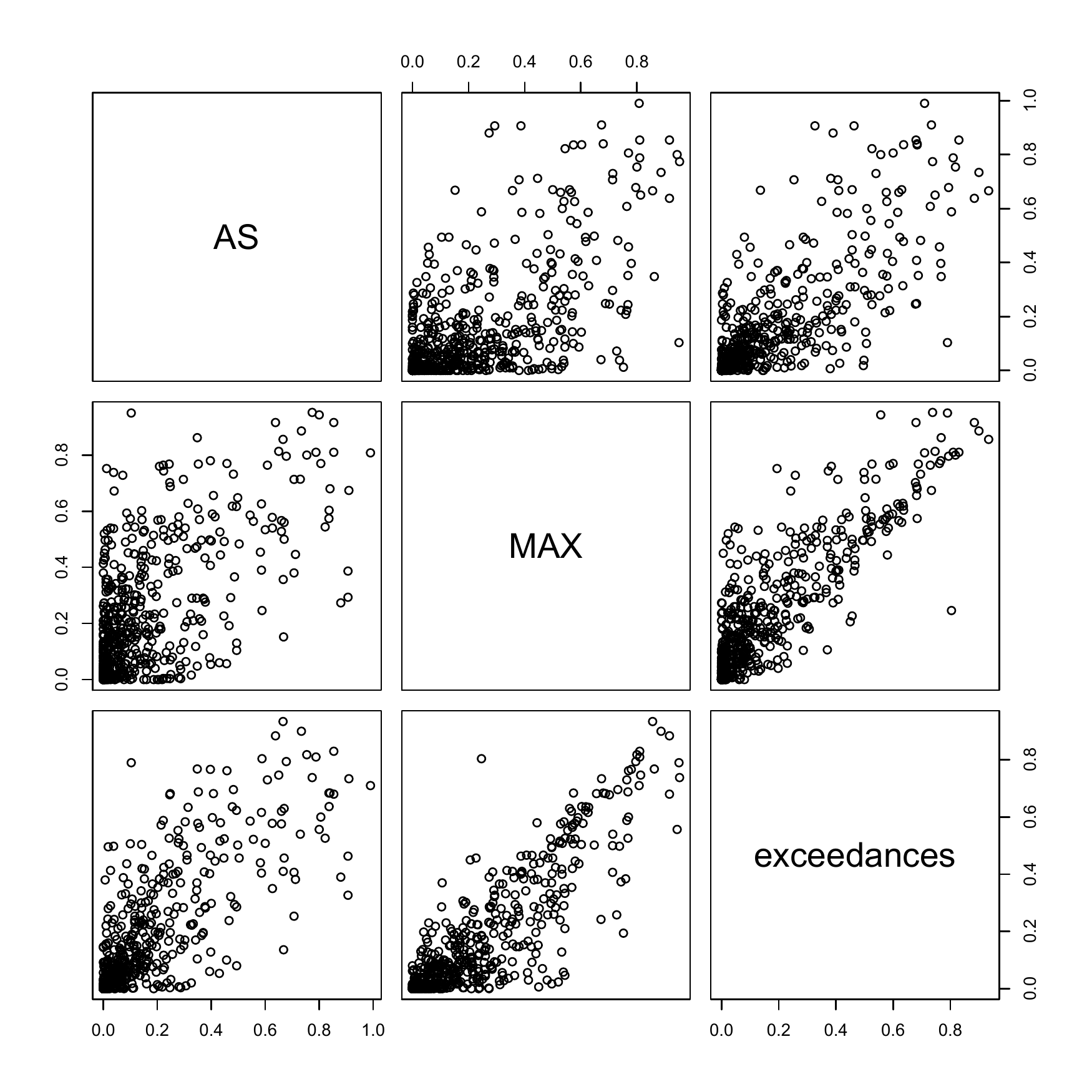}}&
    \subfloat[Venn's diagram]{\includegraphics[scale=0.38]{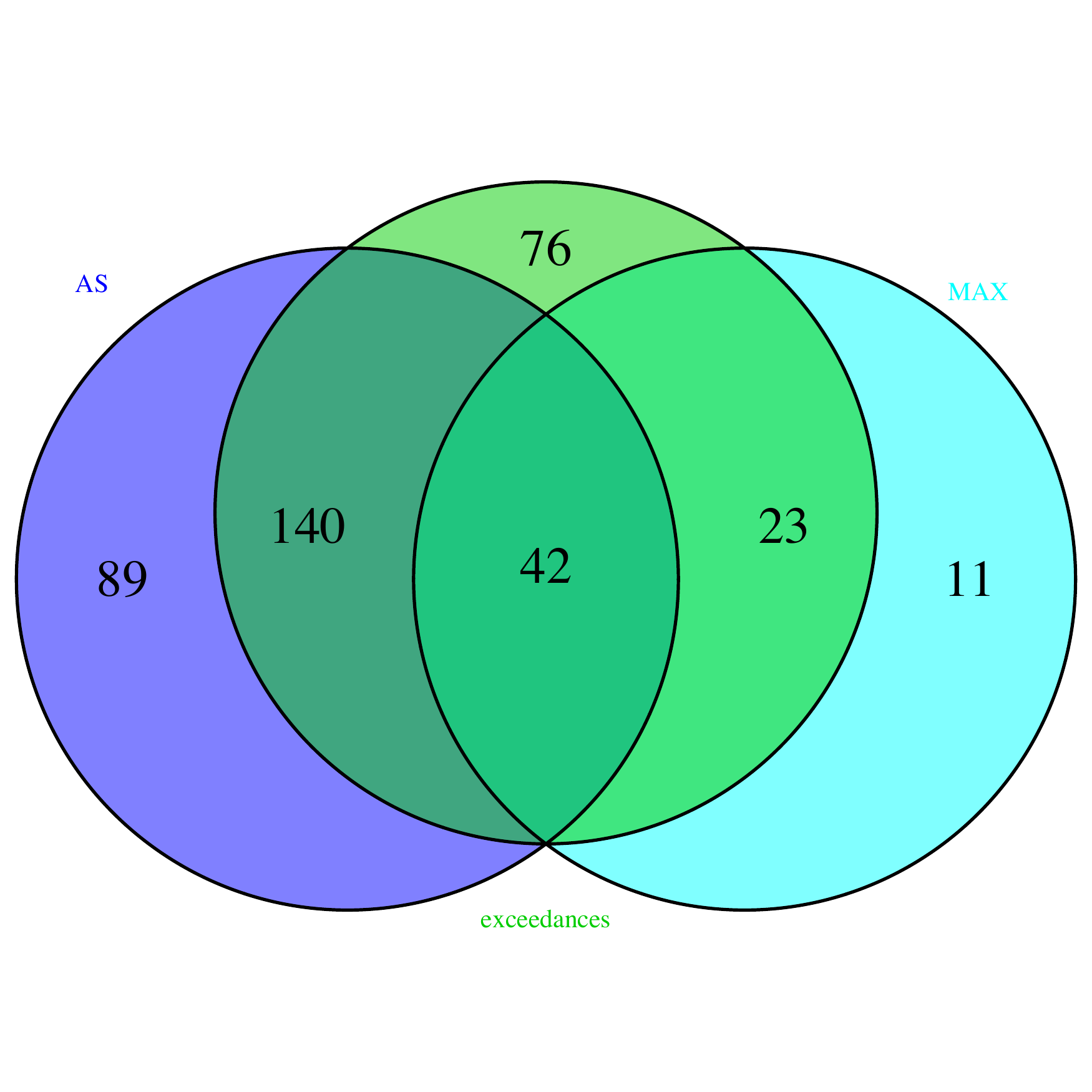}}
\end{tabular}
 \caption{\footnotesize{ P-values for  equality of correlations on 1,320 pathway lists of genes.\ Venn's diagram shows the number of rejected lists with an adjusted p-value smaller than 0.05. }}\label{Fpathway}
\end{center}
\end{figure}

In Figure \ref{Fpathway}(a) we present the approximated p-values using the three dependence-correction tests. In the sum of exceedances test we give the results for $w=0$, although they are quite similar to the p-values found for $w=1$.\ The $18\%$ of the average of squares test p-values, the $9\%$ of the extreme value test p-values and the $19\%$ of the sum of exceedances test p-values are smaller than $0.01$ and under $H_0$ we were expecting only $1\%$.\ About $4\%$ of the lists have the three tests with p-values smaller than $0.01$.\ Moreover, about $35\%$ of the lists have the three p-values larger than $0.10$, indicating some similarity in the correlation matrices even with conditions as different as cancer and healthy.\ We further adjust the p-values for multiple testing by controlling the false discovery rate, and in Figure \ref{Fpathway}(b) we present a Venn's diagram of the adjusted p-values smaller than 0.05.\ Moreover, in Table \ref{TabList} we highlight some of the pathways lists that had significant adjusted p-values in the three tests.

We preferred to use the dependence-correction tests rather than the asymptotic independence ones since the dependence in the sample correlation matrices for each of the lists resulted to be quite strong.\ Besides, the  obtained results assuming  non-parametric distributions were very similar to dependence-correction, and for simplicity we only show the p-values for the latter.

%For instance, using the statistic $T_I$ defined in (\ref{AsymOrNot}) we obtain that less than $2\%$ of all the lists have a value smaller than $1$.
The sample correlations between test p-values are also remarkably large, $0.59$ between average of squares and extreme value, $0.87$ between extreme value and exceedance and $0.75$ between average of squares and exceedance (these are more similar to the values obtained in the simulation study from the dense scenario described in Section \ref{denseCor}  than to the sparse one in Section \ref{powerLaw}).

\begin{table}
\caption{\label{TabList}Lists with p-values smaller than $0.0003$ for all the tests. Highly overlap label corresponds to
pathways lists that contain more than $50\%$ of their genes common to another list.} % title of \footnotesize
\footnotesize
\begin{tabular}{ >{\collectcell\myverb}l<{\endcollectcell} l }
%\hline
&\\
 1- "KEGG_SPLICEOSOME"                                   	&\\		
 2- "KEGG_JAK_STAT_SIGNALING_PATHWAY"      &\\
 3- "BIOCARTA_INFLAM_PATHWAY"   &  (highly overlaps with [2])                 \\
 4- "BIOCARTA_ERYTH_PATHWAY"     & (highly overlaps with [3])               \\
 5- "BIOCARTA_STEM_PATHWAY"      & (highly overlaps with [2] and [3])  \\
 6- "REACTOME_SIGNALING_BY_GPCR"&\\
 7- "REACTOME_GPCR_DOWNSTREAM_SIGNALING"                                                  &\\
 8- "REACTOME_SIGNALING_BY_ILS"&\\
 9- "REACTOME_CYTOKINE_SIGNALING_IMMUNE_SYSTEM"                                                              &\\
10- "REACTOME_TELOMERE_MAINTENANCE"& \\
%\hline
\end{tabular}
%\label{TabList}
%\normalsize
\end{table}

%%%%%%%%%%%%%%%%%%%%%%%%%%%%%%%%%%%%%%%%%%%%%%%%%%%%%%%%%%%%
%%%%%%%%%%%%%%%%%%%%%%%%%%%%%%%%%%%%%%%%%%%%%%%%%%%%%%%%%%%%

\section{Discussion}
In this article we propose three tests for equality of two correlation matrices: average of squares, extreme value and sum of exceedances tests. These are especially useful for high-dimensional dependent datasets.\ We further suggest considering dependence-correction or non-parametric tests instead of asymptotic linear independence  tests when the correlation matrices are known to be dense.\  Asymptotic tests, which assume independence among sample correlation coefficients, are much faster than the other two tests and could be used for highly sparse correlation matrices to speed up the process.\ For dense correlation matrices though, asymptotic tests can produce a non-negligible bias in the approximated p-values when the null hypothesis is true.

The idea of dependence-correction tests diverges with the methods seen so far in the literature.\ For instance, the extreme value test proposed in this paper contrasts with the results by \cite{TonyCai2014} who test  the equality of mean vectors by employing the maximum of the square value of element-wise differences. The authors, as we have also done in Section \ref{GumbelApprox} of the appendix, prove that the limiting distribution of the maximum of dependent samples converges to the extreme value distribution of type I under very mild conditions and they examine this limiting distribution to assess the evidence of the test. We estimate the parameters given permuted samples since its known that the convergence of the parameters to the asymptotic ones is slow and we account for bias that arise in dependent datasets due to estimating correlation of sample correlation coefficients \citep{Olkin1990}.

In terms of test power, for a sensible selection of the exceedance threshold, sum of exceedances test is shown to be the most powerful test for sparse alternatives.\ If the sparsity levels are high,  the extreme value also provides competitive results.\ In contrast, for dense alternatives and small sample size, the average of squares dominates the asymptotic power.\

We use 1,320 pathway lists to test equality of gene dependence's structures between normal and cancer human samples in groups of genes that are known to interact together in a cell.\ A large part of the total number of lists has significantly small p-values.\ Especially, this happens in the average of squares and sum of exceedances tests.\ The extreme value test also gives smaller p-values that expected under the null hypothesis but it is more inclined to not reject $H_0$ than the other two tests.\ This could be an indication, if $H_1$ is true, that dependence structures are closer to the dense alternative scenario rather than the  sparse scenario.

As future work, we intend to use the sum of exceedances test statistic for higher criticism testing \citep{Donoho2004} as a way to avoid the threshold selection problem and maximize the power of the test.

\section*{Acknowledgements}
Adria Caballe Mestres and Claus Mayer acknowledge financial support from the Scottish Government's Rural and Environment Science and Analytical Services Division (RESAS). Natalia Bochkina and Ioannis Papastathopoulos  are grateful to the Alan Turing Institute for the financial support under the EPSRC grant EP/N510129/1.

\section{Appendix}
\subsection{Variance of mean of squares for dependent samples}\label{proof1}
Here we proof the result in Lemma 1 that gives the expression of the variance of the average of squares for dependent random variables. Consider  $n$ dependent random variables $Z = (z_1,\ldots, z_n)$ which marginally follow a standard normal distribution. Take $\mathbf{E}[z_i^2]=\mu_2=1$ and  $\mathbf{E}[z_i^4]=\mu_4=3$ for any $z_i \in Z$ and $\bar{\gamma_2}=2(n(n-1))^{-1} \sum_{i<j}\text{cov}(z_i^2,z_j^2)$ which is function of the dependence structure between variables.

The mean square of elements in $Z$ is found by $S^2 = n^{-1} \sum_{i=1}^n z_i^2$ and has variance  $\text{var}[S^2] = \mathbf{E}[S^4] - \mathbf{E}[S^2]^2$. The second term is determined by $\mu_2$ such that $\mathbf{E}[S^2]^2 = \mu_2^2$. Moreover, the first term is expressed as
$$
\mathbf{E}[S^4] = \mathbf{E}[n^{-2} (\sum_{i=1}^n z_i^2)^2] = \mu_4/n + (\bar{\gamma}_2 + \mu_2)(n-1)/n.
$$
Hence, $\text{var}[S^2] = (\mu_4 - \mu_2^2)/n +  \bar{\gamma}_2 (n-1)/n$.

%\newpage

\subsection{First and second order statistics for estimated exceedances}\label{proof3}
We show the expected value and variance of $(|\hat{d}_t|-w_uu)^2|\hat{d}_t^2>u^2$ for a general case of $d_t$ being any value. This is used in the paper to obtain the lower bound of the power of the sum of exceedances test, and also to select the threshold $u$.

\subsubsection{ Scenario $w_u=0$}
Take $x_t = \hat{d}_t \sim N(d_t, 1)$. Expected value is determined by
\begin{equation} \label{expH1}
\begin{split}
E[x^2_t \mid  x^2_t > u^2] &= \frac{\int_{u}^{\infty} x_t^2(2\pi)^{-1/2} e^{-\frac{(x_t-d_t)^2}{2}} dx_t + \int_{-\infty}^{-u} x_t^2(2\pi)^{-1/2} e^{-\frac{(x_t-d_t)^2}{2}} dx_t}{\Phi(d_t-u) + \Phi(-d_t-u)} \\
& = 1 +d_t^2 +\frac{(u-d_t)\varphi(u-d_t)}{\Phi(d_t-u) + \Phi(-d_t-u)} + \frac{(u+d_t)\varphi(-u-d_t)}{\Phi(d_t-u) + \Phi(-d_t-u)}\\
&+ 2d_t\frac{\varphi(u-d_t) -\varphi(-u-d_t) }{\Phi(d_t-u) + \Phi(-d_t-u)}\\
&= 1 +d_t^2 + A + B,
\end{split}
\end{equation}
where $A = u\{\varphi(u-d_t)+\varphi(u+d_t)\}/\{\Phi(d_t-u) + \Phi(-d_t-u)\}$ and  $B = d_t\{\varphi(u-d_t)-\varphi(u+d_t)\}/\{\Phi(d_t-u) + \Phi(-d_t-u)\}$. If $|d_t|>u$, then $E[x^2_t \mid  x^2_t > u^2] \geq d_t^2 +1$. Under $H_0$, where $d_t =0$, $\mu_0   =  1 + u\frac{\varphi(u)}{1-\Phi(u)}$.

The expression for the variance is
\footnotesize
\begin{equation}\label{varH1}
\begin{split}
\text{var}[x^2_t\mid x^2_t>u^2]  &= \frac{(2\pi)^{-1/2} [ \int_{u}^{\infty} x_t^4e^{-\frac{(x_t-d_t)^2}{2}} dx_t + \int_{-\infty}^{-u} x_t^4 e^{-\frac{(x_t-d_t)^2}{2}} dx_t]}{\Phi(d_t-u) + \Phi(-d_t-u)} - E[x^2_t \mid  x^2_t > u^2]^2 \\
&= d_t^4 + d_t^3 D + d_t^2(6+uC) + d_t(u^2+5)D +(u^3+3u)C +3\\
&- E[x^2_t \mid  x^2_t > u^2]^2,
\end{split}
\end{equation}
\normalsize
where $C= \{(\varphi(u+d_t) +\varphi(u-d_t) \}/\{\Phi(d_t-u) + \Phi(-d_t-u)\}$ and $D= \{(\varphi(u+d_t) -\varphi(u-d_t) \}/\{\Phi(d_t-u) + \Phi(-d_t-u)\}$.
Under $H_0$,  $\sigma_0^2 = 3 + (u^3+3u)\frac{\varphi(u)}{1-\Phi(u)} - \mu_0^2$.

%\newpage
\subsubsection{ Scenario $w_u=1$}
Take $x_t = \hat{d}_t \sim N(d_t, 1)$. Expected value is determined by
\footnotesize
\begin{equation} \label{expH11}
\begin{split}
E[(|x|-u)^2_t \mid  x^2_t > u^2] &= \frac{1}{\sqrt{2\pi}}\left[\frac{\int_{u}^{\infty} (x_t-u)^2 e^{-\frac{(x_t-d_t)^2}{2}} dx_t + \int_{-\infty}^{-u} (-x_t-u)^2e^{-\frac{(x_t-d_t)^2}{2}} dx_t}{\Phi(d_t-u) + \Phi(-d_t-u)}\right] \\
& =  E[x^2_t \mid  x^2_t > u^2] + u^2 - 2u\frac{\varphi(d_t-u) +\varphi(-d_t-u) }{\Phi(d_t-u) + \Phi(-d_t-u)} \\
&- 2d_tu\frac{\Phi(d_t-u) -\Phi(-d_t-u) }{\Phi(d_t-u) + \Phi(-d_t-u)}\\
& = 1 +d_t^2 + u^2 +A + B - E,
\end{split}
\end{equation}
\normalsize
where A and B are defined above, and
$$
E= 2u\frac{\varphi(d_t-u) +\varphi(-d_t-u) }{\Phi(d_t-u) + \Phi(-d_t-u)}
- 2d_tu\frac{\Phi(d_t-u) -\Phi(-d_t-u) }{\Phi(d_t-u) + \Phi(-d_t-u)}.
$$
Note that if $|d_t|>u$, then $E[(|x|-u)^2_t \mid  x^2_t > u^2] \geq (|d_t|-u)^2 +1$ can be used as a lower bound. Under $H_0$,
$\mu_1 = (u^2+1) - u\frac{\varphi(u)}{1-\Phi(u)}$.

The expression for the variance is
\footnotesize
\begin{equation}\label{varH11}
\begin{split}
\text{var}[(|x|-u)^2_t\mid x^2_t>u^2] &= \frac{1}{\sqrt{2\pi}}\left[\frac{\int_{u}^{\infty} (x_t-u)^4 e^{-\frac{(x_t-d_t)^2}{2}} dx_t + \int_{-\infty}^{-u} (-x_t-u)^4e^{-\frac{(x_t-d_t)^2}{2}} dx_t}{\Phi(d_t-u) + \Phi(-d_t-u)}\right]\\
 &- E[(|x|-u)^2_t \mid  x^2_t > u^2]^2 \\
 =&E[x^4_t \mid  x^2_t > u^2] + 6uE[x^2_t \mid  x^2_t > u^2] +u^4 + 4u^3(d_tC-D) - F,
\end{split}
\end{equation}
\normalsize
where
\footnotesize
\begin{align*}
F  &= 8uC + 12ud_t^2C + (4ud_t^3 +12d_tu)(\Phi(d_t-u) - \Phi(-d_t-u))/\{\Phi(d_t-u) + \Phi(-d_t-u)\}\\
&+ \frac{4u\{ (u-d_t)^2\varphi(u-d_t) +(u+d_t)\varphi(u+d_t)\} +12d_tu\{(u-d_t)\varphi(u-d_t) - (u+d_t)\varphi(u+d_t) \}}{\Phi(d_t-u) + \Phi(-d_t-u)}.
\end{align*}
\normalsize
Under $H_0$,  $\sigma_0^2 = 3 +u^4+6u^2- (5u +u^3)\frac{\varphi(u)}{1-\Phi(u)} - \mu_1^2$.

\subsection{Gumbel approximation of extreme value test statistic}\label{GumbelApprox}
Let $V_{tj} = \text{cov}(\hat{d}_{t},\hat{d}_{j})$ be the covariance between two elements in the matrix $\hat{D}$. For op $\in \{=,\neq\}$, we define
$$
\nu_t^{\text{op}} = \sum_{j \in A} I(V_{tj}\,\, \text{op}\,\, 0), \,\,\,\, A= M \setminus \{t\},
$$
so $\nu_t^{=} +\nu_t^{\neq} = m-1$. Following sparsity constrains in  \cite{Meinshausen2006}, the sparsity level $\nu_t^{\neq}$ is assumed to be
$$
\nu_t^{\neq} = O(m^\eta_t) = L(m)\,m^{\eta_t},
$$
where $0\leq \eta_t <1$ and $L(m)$ is a slowly varying function, i.e.,
$\lim\limits_{m\to \infty} L(mx)/L(m) \to 1$.\ Moreover,
$$
\nu_t^{=} = m-1 - O(m^\eta_t) =m\,(1-m^{-1}- L(m)\,m^{\eta_t-1}) = m\,(1+o(1)) = L(m)\,m.
$$
Assume that $\max_{i<j}|V_{tj}|  <1$ and that there exists a
permutation $\hat{D}^{*}$ of elements in $\hat{D}$ such that
$V^* = [\text{cov}(\hat{d}_{t}^*,\hat{d}_{j}^*)]$ is block diagonal. Then for all rows in
$V^{*}$ there exists $h$ such that for all
$j>h\, :\, V_{tj}^* =0$. Let $\epsilon_n \in o(1/\log n)$ and take
$\epsilon$ any positive number such that
$\max_{i<j}|V_{ij}^*|+\epsilon < 1$. Define
$$
\rho_n =
\begin{cases} \max_{t<j}|V_{tj}|+\epsilon, & n<|j-t|\\
  \epsilon_n,& n\geq |k-t|.
\end{cases}
$$
It then follows that $|V_{tj}^*| < \rho_{|j-t|}$, and
$\rho_n\log n \to 0$ as $n\to \infty$. This is a sufficient condition
\citep{Leadbetter1983} for the distribution of
$T_{MAX} = \max\limits_{t \in M} |\hat{d}_{t}|$ to converge weakly to
a Gumbel distribution. % to converge in distribution to a Gumbel.

\subsection{Sub-asymptotic model for structured non-stationary processes}\label{thetaStruct}
 The heuristic approach proposed in this section follows results and notation from \cite{Aldous1992}.
 Let $\mathcal{S}_x = \{t\in M: |\hat{d}_t| \geq x\}$ be  a random set that, for large $x$, defines a sparse
 mosaic on the sub-integer lattice $\mathbb{Z}^2$ corresponding to the lower triangular matrix $M$ (defined in~eq.\eqref{standDif2}).\
 We assume a structured dependence structure on the process $(\hat{d}_t: t\in M)$ such that
 $\mathcal{S}_x$ contains several (near) independent clusters defined by a compound Bernoulli process
 with cluster intensity $\lambda_x(t)$. \ Let $C_x(t)$ denote the cluster area (or cardinality) at point $t$,
 and assume that as the number of variables increase, $C_x(t)$, in any position $t\in M$,  is finite
 and does not exceed a given constant $\kappa$.  Besides,  assume that $\lambda_x(t)$ and $C_x(t)$
 do not vary much as $t$ moves around the same cluster. For  $x(m)=\mu(m)+\sigma(m)\,x$, $x\in\mathbb{R}$,  the distribution
 of $T_{M} = \max_{t \in M} |\hat{d}_t|$ can be approximated by
 \begin{IEEEeqnarray*}{rCl}
   \Pr(T_{M} < x(m)) &=& \Pr(\mathcal{S}_{x(m)} \cap M\,\, \text{empty}) \\
   &\doteq & \exp\left(-\int_M \lambda_{x(m)}(t)~\mathrm{d}t\right)
   % &\doteq& \exp\left(-\lambda_x_m \, |A|\right) \\
\\
  & \doteq & \exp\left\{-\int_M \frac{\Pr(|\hat{d}| > x(m))}{\mathbb{E}(C_t^{x(m)})}~\mathrm{d}t \right\}\\
   &=& \exp\left\{-\Pr(|\hat{d}| > x(m)) \int_M \frac{1}{\mathbb{E}(C_t^{x(m)})}~\mathrm{d}t \right\}\\
    &=& \exp\left\{-\Pr(|\hat{d}| > x(m))  \sum_{t \in M} \frac{1}{\mathbb{E}(C_t^{x(m)})}\right\},
 \end{IEEEeqnarray*}
 where $\hat{d}\sim N(0,1)$, $\mathbb{E}(C_t^x)$ is the expected
 cluster area at cell $t$ and threshold level $x$.\
 %Asymptotic theory of extreme values states that  threshold $u \to \sup\{ |\hat{d}_t|: \Phi(|\hat{d}_t|) <1 \}$.\ However,
 The result obtained above  is equivalent to the cumulative distribution function of the cluster maxima for sub-asymptotic models ( $u < \sup\{ |\hat{d}_t|: \Phi(|\hat{d}_t|) <1 \} $)  in a stationary process \citep{Eastoe2012},
 \begin{IEEEeqnarray*}{rCl}\label{faf}
   \Pr(T_{M} < x) &=&  \exp\left\{-m\theta_x\Pr(|\hat{d}_t| > x)\right\}  \\
   &\doteq&  \exp\left[- m p_u \theta_x \exp\{- (x - u)/\sigma_u\}\right], \mbox{\hspace{0.3cm}} (x\geq u)
 \end{IEEEeqnarray*}
 when $m \theta_x =  \sum_{t=1}^{m} \frac{1}{\mathbb{E}(C_t^x)}$ and with $p_u = \Pr(|\hat{d}_t| > u)$.

% Note that $\mathbb{E}(C_t^x)$ depends on the exceedance level $x$ and position of the cell $t$.\ Since this is a Gaussian process with long range independence, we know that $\mathbb{E}(C_t^x)\to 1$ and also $\theta_x\to 1$ as $x\to \infty$, thus

%Nevertheless, for sub-asymptotic models, $\mathbb{E}(C_t^x)$ is larger than one for some $t, x$, with $\theta_x<1$ and alternative approximations may represent $\Pr(T_{M} < x)$ better than the expression given in eq.\ \eqref{pvalMax}.\ From max-stability property of the Gumbel distribution, it follows that if  $H(x) = \{G(x)\}^\theta$ where $G(x)$ is a Gumbel$(\mu, \sigma)$, then $H(x)$ is also a Gumbel with different parameters $(\mu_\theta, \sigma_\theta)$ such that

\subsection{Asymptotic power}\label{proof5}
Let's first acknowledge the Mill's ratio which approximates $\Phi(-x) \doteq \frac{\varphi(x)}{x}$, where $\varphi(x) = e^{-\frac{1}{2} x^2}$, when $x$ is  large. We recall that we use the set of variables  $(\hat{d}_t: t\in M)$, with $m = \text{card}(M)$ such that $\mathcal{S}_d = \{ t\in M: d_{t} \neq 0\}$ and $s= \text{Card}(\mathcal{S}_d )$ is the sparsity level.  We assume that  $|g(r_{Y_{t}}) - g(r_{X_{t}})| = \delta_t $ for all $t\in \mathcal{S}_d$ with $d_t = \sqrt{n-3}\delta_t$. Moreover, we consider normality for the  Fisher transform correlation differences such that for all $t \in \mathcal{S}_d$,   $\hat{d}_t \sim N(\delta_t,(n-3)^{-1})$ and for all $t \not\in \mathcal{S}_d$,   $\hat{d}_t \sim N(0,(n-3)^{-1})$.

The power of the test is given by the probability of rejecting the null hypothesis when the $H_1$ is true. Hence, the objective is to find the test that provides the maximum power. For all tests ($q = s, m, e$), we  define a rejecting level $t_{q,\alpha}$ such that we reject the null hypothesis when the observed test statistic is larger than $t_{q,\alpha}$ at significance level $\alpha$.

\subsubsection{Asymptotic power for average of squares test}\label{asymSumSquaresP}
Here we assume that the test statistic $T_{S}$ defined in eq.~\ref{TestStatistics} %(6)
of  the main paper is well approximated by a normal distribution under both  $H_0$ and $H_1$. We define $\mu_{H_0}$ and $\sigma_{H_0}^2$ as the expected value and variance of $T_{S}$ when $H_0$ holds.
Moreover, $\mu_{H_1}$ and $\sigma_{H_1}^2$ are the correspondent expected value and variance of $T_{S}$ when $H_1$ holds.
The power of the average of squares test is
\begin{equation}\label{eq:powerAS1}
\Pr(T_{S} \geq t_{S,\alpha} \mid  H_1) \doteq \Pr\left( Z \geq \frac{\mu_{H_1}  - t_{S,\alpha}}{\sqrt{\sigma_{H_1}^2}}\right),
\end{equation}
approximated using the Mill's ratio, with rejecting level  given by $t_{S,\alpha} = \mu_{H_0} + z_{\alpha}\sqrt{\sigma_{H_0}^2}$.

Denote  $\delta_0^2 = \sum_{t\in \mathcal{S}_d} \delta_t^2 $ and recall that $\bar{\gamma}_2 = 2(m^2 - m)^{-1}\sum_{t<h} \text{cov}(\hat{d}_{t}^2,\hat{d}_{h}^2\mid H_0)$. Under $H_0$, the parameters  $\mu_{H_0} \doteq 1$ and  $\sigma^2_{H_0} \doteq \frac{2}{m} \{1+ (m-1)\bar{\gamma}_2/2\}$. The expected value of $T_{S}$ under $H_1$ is found by a weighted average $\mu_{H_1} = (m-s)\mu_0/m + s\mu_1/m$ with $\mu_0 = \mathbf{E}[\hat{d}_{t}^2 \mid  t \not\in \mathcal{S}_d ]  \doteq 1$ and $\mu_1 = \text{var}[\hat{d}_{t} \mid   t \in \mathcal{S}_d ] + \mathbf{E}[\hat{d}_{t} \mid  t \in \mathcal{S}_d]^2 \doteq 1 + d_t^2$.
Similarly, the parameter $\sigma_{H_1}^2$ can be found by the variance of a weighted average, so
$\sigma_{H_1}^2  = 2/m (1 + 2s(n-3)\delta_0^2/m +  (m-1)\bar{\gamma}_2'/2)$ where
$\bar{\gamma}_2' = 2(m^2 - m)^{-1}\sum_{t<h} \text{cov}(\hat{d}_{t}^2,\hat{d}_{h}^2\mid H_1)$.
Note that $\bar{\gamma}_2'$ is different to $\bar{\gamma}_2$ as it depends on the values $(d_t, t\in \mathcal{S}_d)$.
 Plugging in the expressions for $t_{S,\alpha}$, $\mu_{H_1}$ and $\sigma_{H_1}^2$ in \eqref{eq:powerAS1}, we obtain the stated expression for the power.

%powermaxtestproof

\subsubsection{Asymptotic power of the extreme value test}\label{asymMaxP}
We assume $(\hat{d}_t)\sim MVN$, $t\in M$ under both  $H_0$ and $H_1$. Hence, the maximum $T_{M} = \max_{t \in M} |\hat{d}_{t}|$, in the limit, is well represented by a Gumbel distribution.
We further define the parameters $\mu_t = \mathbf{E}[\hat{d}_{t} \mid  t \in \mathcal{S}_d ]$, $\sigma_t^2 = \text{var}[\hat{d}_{t} \mid  t \in \mathcal{S}_d]$ with $|\mu_t|$ being sufficiently large. Assume independence on the sequence $(\hat{d}_t)$, the power of the extreme value test is defined by
\begin{align*}
\Pr(T_{M} \geq t_{M,\alpha} \mid  H_1) &= 1 -   \Pr( |d_t| <  t_{M,\alpha}, \,\, \forall t) \geq 1 -   \Pr( |d_t| <  t_{M,\alpha} \, \colon\, t \in \mathcal{S}_d)\\
&= 1 -   \Pr\left(  \frac{ -t_{M,\alpha} - \mu_t}{\sigma_t}< Z_t < \frac{ t_{M,\alpha} - \mu_t}{\sigma_t}, \, t \in \mathcal{S}_d\right) \\
&\geq 1 -  \Pr\left( Z_t < \frac{ t_{M,\alpha} - |\mu_t|}{\sigma_t}, \, t \in \mathcal{S}_d\right),
\end{align*}
where $Z_t = (|d_t| - \mu_t)/\sigma_t$.
The rejecting level $t_{M,\alpha}$ is found using the quantile function of the Gumbel distribution that in the limit ascertains that
$$
Q_G(\alpha) \doteq (2\log 2m)^{1/2} - \frac{\log \log 2m + \log (4\pi \log_2 2)}{2(2\log2m)^{1/2}} - \frac{\log(-\log(\alpha))}{(2\log2m)^{1/2}}.
$$
We use the main term of the expression to find  $Q_G(\alpha)$ such that
$$
t_{M,\alpha} = (2\log 2m)^{1/2} - \frac{\log(-\log(\alpha))}{(2\log2m)^{1/2}}  > Q_G(\alpha).
$$
For the expected value of the test statistic under $H_1$ we use $|\mu_t| \doteq  \delta_t\sqrt{n-3}$, and  for the variance we approximate $\sigma_t^2  \doteq \text{var}(\hat{d}_{t}) \doteq 1$, for all $t\in \mathcal{S}_d$.
%
%[Note that for large $n$,  all  coefficients $\hat{d}_{t}$ will have the same sign as $d_{t}$, and thus the approximations will be exact.]
%

If $s=|\mathcal{S}_d| \to \infty$ and the conditions of the Gumbel approximation described in Section~\ref{GumbelApprox} hold (namely that the maximum correlation between pairs of $d_t$, $t\in \mathcal{S}_d$, is bounded above by a constant strictly less than 1), we have
\footnotesize
\begin{align*}
\Pr(T_{M} \geq t_{M,\alpha} \mid  H_1)
%&\geq 1 -  \exp\{ -\exp\{ -[(t_{m,\alpha} - (n-3)^{1/2}\delta_t - (2\log 2 s)^{1/2})(2\log 2s)^{1/2} ] \}\}\\
%&\geq 1 -  \exp\{ -\exp\{- \log(-\log(\alpha))(\log s/ \log m)^{1/2} + (\mu_m-\mu_s)/\sigma_s- (n-3)^{1/2}\delta_t (2\log2s)^{1/2} \}\}\\
% &= 1 -  \exp\{ -\exp\{   - (2\log2s)^{1/2}[(n-3)^{1/2}\delta_t -(2\log 2m)^{1/2} + (2\log 2s)^{1/2} + \log(-\log(\alpha))/(2 \log 2m)^{1/2}] \}\}\\
&\geq 1 -  \Pr\left( Z_t < \frac{ t_{M,\alpha} - \min_{t \in \mathcal{S}_d}|\mu_t|}{\sigma_t}, \, t \in \mathcal{S}_d\right)\\
 &\geq 1 -  \exp\{ -\exp\{   - (2\log2s)^{1/2}[(n-3)^{1/2}\min_{t \in \mathcal{S}_d}\delta_t -(2\log 2m)^{1/2} + (2\log 2s)^{1/2}] \}\}\\
  &\approx 1 -  \exp\{ -\exp\{   - (2\log2s)^{1/2}[(n-3)^{1/2}\min_{t \in \mathcal{S}_d}\delta_t -(2\log 2m)^{1/2}] \}\}.
 \end{align*}
\normalsize

% t_{m,\alpha} = mu_m +\sigma_m q_\alpha
%\Pr( [T_{MAX} - (n-3)^{1/2}\delta_t - \mu_m)/\sigma_m\geq (t_{m,\alpha}- (n-3)^{1/2}\delta_t- \mu_m)/\sigma_m \mid  H_1)
%&\geq 1 -  \exp\{ -\exp\{ -[(t_{m,\alpha} - (n-3)^{1/2}\delta_t)(2\log2m)^{-1/2} +(2\log 2m)^{1/2}], \}\},

%(t_{m,\alpha}- (n-3)^{1/2}\delta_t- \mu_s)/\sigma_s =  q_\alpha \sigma_m/\sigma_s + (\mu_m-\mu_s)/\sigma_s- (n-3)^{1/2}\delta_t/\sigma_s
% here \mu_m = (2\log 2m)^{1/2}, \sigma_m = (2\log2m)^{-1/2}

% (\mu_m-\mu_s)/\sigma_s = 4[(\log 2m)^{1/2} - (\log 2s)^{1/2}](\log 2s)^{1/2} = 4[(\log 2m \log 2s)^{1/2} - \log 2s

% \exp\{- a\log(-\log(\alpha)) }=\exp\{\log([-\log(\alpha)]^{-a}) }=[-\log(\alpha)]^{-a}

If $s=|\mathcal{S}_d|$ is a constant, then, using the Mill's ratio to approximate the normal probabilities,
\footnotesize
\begin{align*}
\Pr(T_{M} \geq t_{M,\alpha} \mid  H_1)
&\geq 1 -   \Pr\left( Z_t < \frac{ t_{M,\alpha} - |\mu_t|}{\sigma_t}, \, t \in \mathcal{S}_d\right)
\geq 1 -  \min_{t \in \mathcal{S}_d}\Pr\left( Z_t < \frac{ t_{M,\alpha} - |\mu_t|}{\sigma_t} \right)\\
& \geq 1 -   \min_{t \in \mathcal{S}_d}  \exp\left[  -\frac{1}{2} \left\{ (n-3)^{1/2}\delta_t - \left( (2\log 2m)^{1/2} - \frac{\log(-\log(\alpha))}{(2\log2m)^{1/2}}\right)\right\}^2\right]\\
& \approx 1 -   \min_{t \in \mathcal{S}_d}  \exp\left[  -\frac{1}{2} \left\{ (n-3)^{1/2}\delta_t -  (2\log 2m)^{1/2}\right\}^2\right].
\end{align*}
\normalsize

%where extremal index $\theta_m$ tends to $1$ as $m \to \infty$.
%

\subsubsection{Asymptotic power of the exceedances test}\label{asymExceP1}
We set an arbitrary large threshold $u$, such that we define set $\mathcal{S}_u = \{ t \in M: |\hat{d}_{t}| >u\}$.
We define the probabilities $\eta_0 = \Pr(t\in \mathcal{S}_u  \mid  t \not\in \mathcal{S}_d)$ and $\eta_t = \Pr(t\in \mathcal{S}_u \mid  t \in \mathcal{S}_d)$ as well as the standard normal distribution density function at quantile $u$ which we denote by $\varphi(u)$.
Under both $H_0$ and $H_1$, we approximate the test statistic $T_E^{(w)}$ described in eq.~\eqref{TestStatistics} by a normal distribution. We define $\mu_{H_0}(m,w)$ and $\sigma_{H_0}^2(m,w)$ as the expected value and variance of $T_{E}^{(w)}$ when $H_0$ holds.
Moreover, $\mu_{H_1}(m,w)$ and $\sigma_{H_1}^2(m,w)$ are the correspondent expected value and variance of $T_{E}^{(w)}$ when $H_1$ holds.
To find both $\mu_{H_1}(m,w)$ and $\sigma_{H_1}^2(m,w)$, we redefine the measures in eq.\eqref{sumEXCparam} by assuming that the expected value of
$\hat{d_t}$ can be different from zero for some $t\in M$:
\begin{IEEEeqnarray}{rCl}
\gamma_{u_{tj}}^{(H1,w)} &=& \text{cov}((|\hat{d}_t|-uw)^2,(|\hat{d}_j|-uw)^2 \mid  \hat{d}_t^2>u,\hat{d}_j^2>u), \nonumber\\
\eta_t  &=& \Pr( |\hat{d}_{t}| > u),\nonumber\\
\phi_{tj}^{H1} &=& \Pr(\hat{d}_t^2>u^2,\hat{d}_j^2>u^2),\nonumber
\end{IEEEeqnarray}
The power is described by
\begin{equation*}
\Pr(T_{E}^{(w)} \geq t_{E,\alpha}^{(w)}\mid  H_1) \doteq \Pr\left( Z \geq \frac{\mu_{H_1}(m,w)  - t_{E,\alpha}^{(w)}}{\sqrt{\sigma^2_{H_1}(m,w)}}\right),
\end{equation*}
where $\mu_{H_1}^{(w)} =  (m-s) \eta_0 \mu_w + \sum_{t\in \mathcal{S}_d} \eta_t \mu_{t_w}$,   rejecting level  $t_{E,\alpha}^{(w)}= \mu_{H_0}^{(w)} + z_{\alpha}\sqrt{\sigma^2_{H_0}(m,w)}$, and
\begin{IEEEeqnarray}{rCl}
\sigma^2_{H_1}(m,w) &=& \sum_{t\in \mathcal{S}_d} \eta_{t}\{(1-\eta_{t})\,\mu_{t_w}^2 + \sigma_{t_w}^2\} +
(m-s) \eta_{0}\{(1-\eta_{0})\,\mu_{w}^2 + \sigma_{w}^2\} + C_w, \nonumber
\end{IEEEeqnarray}
where $C_w = \sum_{t,h \in M, t\neq h} (\gamma_{u_{th}}^{(H1,w)} +\mu_{t_w}\mu_{h_w}) \phi_{th}^{H1}- \eta_{t}\mu_{t_w}\eta_{h_w}\mu_{h_w}$ is different from zero if elements in $\hat{D}^2$ are dependent.
Let $\mu_{H_0}(m,w)  = \mu(m,w)$ and $\sigma^2_{H_0}(m,w) = \sigma^2(m,w)$ defined by eq.~\eqref{musigma3}.  The lower bound for the asymptotic power of sum of exceedances test, with $w = \{0, 1\}$, is
\begin{align*}
&\Pr(T_{E}^{(w)} \geq t_{E,\alpha}^{(w)} \mid  H_1) \geq 1 -
  \exp\left\{ -\frac{1}{2}\left(\frac{\sum_{t\in \mathcal{S}_d}  \mu_{t_w} \eta_t - s\,\eta_0\,\mu_w -z_\alpha\,\sigma_{H_0}(m,w)}{
\sigma_{H_1}(m,w) } \right)^2\right\}.
\end{align*}

Let $\mathcal{S}_{du} = \{t\in M, |d_t|\gg u\}$ with $s_u = |\mathcal{S}_{du}|$. For $w=0$, when $(n,m,u)\to \infty$,  under weak independence,
i.e., $C_w \ll \sigma^2_{H_1}(m,w)$, the asymptotic power leading terms ares
$$
\frac{\sum_{t\in \mathcal{S}_{du}}{d_t^2} -  B_0 (s\eta_0^{1/2}  + z_\alpha (2m)^{1/2})}{\sqrt{\sum_{t\in \mathcal{S}_{du}}{d_t^2}   + mB_0^2}},
$$
where $B_0 = u^2\eta_0^{1/2}$. Let $\delta_{00}^2 = s_u^{-1}\sum_{t\in \mathcal{S}_{du}}{d_t^2}$, asymptotic recovery condition is
$$
\delta_{00}^2 \gg \frac{u^2}{n}\frac{\max(1, s\eta_0, (2m\eta_0)^{1/2})}{s_u},
$$
If $s_u = k \max(1, s\eta_0,(2m\eta_0)^{1/2})$, for any positive integer $k$, and $d_t^2/u^2 \to \infty$, for any $t\in \mathcal{S}_{du}$, $\Pr(T_{E}^{(0)} \geq t_{E,\alpha}^{(0)} \mid  H_1) \to 1$.

Similarly for $w=1$,  when $(n,m,u)\to \infty$, $\mu_1 \approx 2/(u^2-1)$ and $\sigma_1^2 \approx 4/(u^2-1)^2$ (these rates can be found using L'Hospital rule), and  similar weak independence conditions,  the  asymptotic power leading terms are
$$
\frac{\sum_{t\in \mathcal{S}_{du}}{|d_t|-u^2}  -  B_1 (s_u\eta_0^{1/2}  + z_\alpha (2m)^{1/2})}{\sqrt{\sum_{t\in \mathcal{S}_{du}}{|d_t|-u^2} + 2mB_1^2}},
$$
where $B_1 = 2\eta_0/(u^2-1)$.  Let $\delta_{01}^2 = s_u^{-1}\sum_{t\in \mathcal{S}_{du}}{(|d_t|- u)^2}$,  asymptotic recovery condition is
$$
\delta_{01}^2 \gg 2/(u^2+1)\frac{\max(1, s\eta_0, (2m\eta_0)^{1/2})}{s_{du}},
$$
If $s_u = k\max(1, s\eta_0,(2m\eta_0)^{1/2})$, for any positive integer $k$, and $d_t^2/u^2 \to \infty$, for any $t\in \mathcal{S}_{du}$,  $\Pr(T_{E}^{(1)} \geq t_{E,\alpha}^{(1)} \mid  H_1) \to 1$.

\bibliographystyle{rss}

%\include{referencies}

%%%%%%%%%%%%%%%%%%%%%%%%%%%%%%%%%%%%%%%%%%%%%%%%%%%%%%%%%%%%
%%%%%%%%%%%%%%%%%%%%%%%%%%%%%%%%%%%%%%%%%%%%%%%%%%%%%%%%%%%%

\end{document}